\documentclass[usenatbib]{mn2e}

\usepackage{graphicx}
\usepackage{color}
\usepackage{hyperref}

\graphicspath{{Figures/}}

\def\simgt{\hbox{\rlap{\raise 0.425ex\hbox{$>$}}\lower 0.65ex\hbox{$\sim$}}}
\def\simlt{\hbox{\rlap{\raise 0.425ex\hbox{$<$}}\lower 0.65ex\hbox{$\sim$}}}
\def\arcsec{^{\prime\prime}}
\def\arcmin{^\prime}

\def \nii {[N{\small~II}]}

\def \caii {Ca{\small~II}}

\def \ha {H$\alpha$}
\def \hb {H$\beta$}

\def \twodfdr {2{\sc dfdr}}

\title[The SAMI Kinematic Morphology-Density Relation.]{The SAMI Pilot Survey: The Kinematic Morphology-Density Relation in Abell 85, Abell 168 and Abell 2399}

\author[Fogarty et al.,]
{L. M. R. Fogarty$^{1,2}$\thanks{l.fogarty@physics.usyd.edu.au}, Nicholas Scott$^{1,2}$, Matt S. Owers$^{3}$, S. Brough$^{3}$, Scott M. Croom$^{1,2}$, 
\newauthor Michael B. Pracy$^{1}$, R. C. W. Houghton$^{5}$, Joss Bland-Hawthorn$^{1}$, Matthew Colless$^{4}$, 
\newauthor Roger L. Davies$^{5}$, D. Heath Jones$^{6}$, J. T. Allen$^{1,2}$, Julia J. Bryant$^{1,2,3}$,  
\newauthor Michael Goodwin$^{3}$, Andrew W. Green$^{3}$, Iraklis S. Konstantopoulos$^{3}$, J.S. Lawrence$^{3}$, 
\newauthor Samuel Richards$^{1,2,3}$, Luca Cortese$^{7}$, Rob Sharp$^{4}$.
\\
$^{1}$ Sydney Institute for Astronomy, School of Physics, University of Sydney, NSW 2006, Australia.\\ 
$^{2}$ ARC Centre of Excellence for All-Sky Astrophysics (CAASTRO).\\
$^{3}$ Australian Astronomical Observatory, PO Box 915, North Ryde, NSW 1670, Australia.\\
$^{4}$ Research School of Astronomy and Astrophysics, Australian National University, Canberra ACT 2611, Australia. \\
$^{5}$ Astrophysics, Department of Physics, University of Oxford, Denys Wilkinson Building, Keble Rd., Oxford, OX1 3RH, UK. \\
$^{6}$ School of Physics, Monash University, Clayton, VIC 3800, Australia. \\
$^{7}$ Centre for Astrophysics and Supercomputing, Swinburne University of Technology, Hawthorn, VIC 3122, Australia.
}

\begin{document}

\maketitle

\newcommand{\fmmm}[1]{\mbox{$#1$}}
\newcommand{\scnd}{\mbox{\fmmm{''}\hskip-0.3em .}}
\newcommand{\scnp}{\mbox{\fmmm{''}}}

\begin{abstract}

We examine the kinematic morphology of early-type galaxies (ETGs) in three galaxy clusters Abell 85, 168 and 2399. Using data from the Sydney-AAO Multi-object Integral field spectrograph (SAMI) we measured spatially-resolved kinematics for 79 ETGs in these clusters. We calculate $\lambda_{R}$, a proxy for the projected specific stellar angular momentum, for each galaxy and classify the 79 ETGs in our samples as fast or slow rotators. We calculate the fraction of slow rotators in the ETG populations ($f_{SR}$) of the clusters to be $0.21\pm0.08$, $0.08\pm0.08$ and $0.12\pm0.06$ for Abell 85, 168 and 2399 respectively, with an overall fraction of $0.15\pm0.04$. These numbers are broadly consistent with the values found in the literature, confirming recent work asserting that the fraction of slow rotators in the ETG population is constant across many orders of magnitude in global environment.

We examine the distribution of kinematic classes in each cluster as a function of environment using the projected density of galaxies: the kinematic morphology-density relation. We find that in Abell 85 $f_{SR}$ increases in higher density regions but in Abell 168 and Abell 2399 this trend is not seen. We examine the differences between the individual clusters to explain this. In addition, we find slow rotators on the outskirts of two of the clusters studied, Abell 85 and 2399. These galaxies reside in intermediate to low density regions and have clearly not formed at the centre of a cluster environment. We hypothesise that they formed at the centres of groups and are falling into the clusters for the first time.

\end{abstract}

\begin{keywords}
techniques: imaging spectroscopy -- galaxies: kinematics and dynamics -- galaxies: ellipticals -- galaxies: clusters
\end{keywords}

\section{Introduction}
\label{sec:intro}

Early-type galaxies (ETGs) have been studied extensively since the quest to understand galaxy formation began. Exactly what defines an ETG, however, can be subtly different depending on many contextual factors. Here we define an ETG to be a galaxy with a smooth symmetric light profile that does not exhibit spiral structure. Within this definition ETGs are traditionally classified morphologically as elliptical and S0 galaxies. However, this classification does not necessarily reflect physical processes within the galaxies themselves. To understand the formation of ETGs we wish to probe these processes and interpret their impact on ETGs and their formation histories. One way to do this is to probe the angular momentum of ETGs through their measured kinematics and instead classify them based on a {\it kinematic morphology}. The SAURON and ATLAS$^{\rm{3D}}$ surveys introduced this technique \citep{Emsellem2007, Cappellari2007, Emsellem2011} and a new classification system for ETGs. In this system high angular momentum ETGs are classified as as fast rotators (FRs) and low angular momentum ETGs as slow rotators (SRs).

In this new framework most ETGs ($\sim86\%$) are found to exhibit regular rotation, whereas a comparatively smaller fraction of objects show no or weak signs of rotation. Many (50\%) elliptical and most (90\%) S0 galaxies turn out to be FRs. These galaxies are axisymmetric oblate spheroids and are thought to harbour stellar disks \citep{Krajnovic2011}. SRs on the other hand are thought to mostly be true dispersion dominated systems, which may be mildly triaxial \citep{Emsellem2007}. Given the difference in their dynamics, the formation mechanisms responsible for FRs and SRs must be different. However, the dominant factor in the formation of SRs and FRs is still open for debate.

Single merging event models by \citet{Bois2011} show that it is possible to create FRs and SRs in major merger scenarios. However, the creation of a SR by a single major merger relies on the orbits of the two progenitor galaxies being exactly retrograde, and this scenario creates a SR with many signatures of being a double sigma galaxy (i.e. two counter-rotating disks), thus not a true dispersion-dominated system. Investigating a population of galaxies, \citet{Khochfar2011} use semi-analytic models to show that in merging scenarios a FR is the usual product, either through maintaining a small amount of stars in a stellar disk or regrowing the disk after the merger event. Conversely, they find that although SRs are likely to have undergone more major mergers their recent past was dominated by multiple minor mergers with random trajectories. These minor events do not allow a stellar disk to build up in the remnant and can therefore lead to the formation of true SRs. \citet{Naab2013} addressed the formation of FRs and SRs by studying cosmological hydrodynamical simulations of 44 central galaxies and satellites, with the unique ability to link the kinematic features in their $z=0$ galaxies with the formation histories of individual objects. They concluded that it is not possible to single out individual FR and SR formation scenarios, instead finding it possible to make both classes of ETGs by several means. Factors like the mass ratio in merging events, the amount of gas present, and when in the history of a particular galaxy major merger events occured, can all impact the type of ETG formed.

It has been shown in work by \citet{Oemler1974} and \citet{DavisGeller1976} that the relative fraction of different galaxy morphological types varies with environment, such that ETGs are more prevalent in galaxy clusters at the expense of late-type galaxies (LTGs, here including both spiral and irregular galaxies). Using the local galaxy density as an independent variable to study this effect, \citet{Dressler1980} found a very clear trend with density, such that the population fraction of ETGs increased, with a corresponding decrease in the LTG fraction, at higher local density. This is the morphology-density relation. Crucially, this relation was found to hold for all types of clusters studied - high to low mass, regular and irregular, and some strong X-ray sources.

The importance of environment in the formation of FRs and SRs has been investigated in several recent studies. \citet{Cappellari2011}, using the ATLAS$^{3\rm{D}}$ sample of 260 ETGs in a 42\,Mpc volume, showed that the fraction of SRs in their overall parent galaxy sample (including ETGs and LTGs), increases sharply in the centre of the Virgo cluster. Analagous to the morphology-density relation, this is known as the kinematic morphology-density relation. The increase in the fraction of SRs at the centre of the Virgo cluster suggests that SRs may be formed more efficiently in denser environments. \citet{DEugenio2013} followed this work by investigating the kinematic morphology-density relation in Abell 1689, a much higher mass and denser cluster than Virgo. Instead of global galaxy fractions they used the metric of SR fraction, $f_{SR}$, defined as the fraction of SRs in the ETG population. They found a trend of increasing $f_{SR}$ with increasing local density, analagous to the trend in the fraction of SRs in the total population seen in ATLAS$^{3\rm{D}}$. This work was then followed by a study of the Coma cluster by \citet{Houghton2013} who saw the same trend of increasing  $f_{SR}$ with increasing local density, but also noted that the global value for $f_{SR}$ was constant across all samples. That is to say that across many types of global host environment (GHE; such as low mass cluster, high mass cluster, field etc.)  the relative number of SRs and FRs is constant. This implies that the formation mechanism for SRs must be equally efficient across a wide range of GHEs, i.e. SRs are not only formed in clusters. However within a particular GHE, \citet{Houghton2013} did confirm the strong dependence of $f_{SR}$ on local point environment (LPE; defined as the local environment of the galaxy) suggesting that perhaps SRs migrate to the centre of their host clusters through dynamical friction, despite forming elsewhere. \citet{Scott2014} investigated the kinematic morphology-density relation in the low mass Fornax cluster, finding that $f_{SR}$ increases towards the centre of the cluster. The trend seen is weaker in Fornax than the more massive clusters studied (Virgo, Coma and Abell 1689). \citet{Scott2014} also find that even in mass-matched samples of SRs and FRs the SRs are more likely to live in denser LPEs. This implies that dynamical friction alone cannot be responsible for the kinematic morphology-density relation.

The causes and nature of the kinematic morphology-density relation, and its importance in relation to the formation of ETGs, are still not well understood. Outstanding questions include, what is the dominant formation mechanism for SRs and FRs, if any? Are SRs mostly formed in situ at the centre of their GHEs or do they migrate there through mass segregation by dynamical friction, or both? What roles do the processes of major and minor merging play in the creation of FRs and SRs and their observed distributions? These questions can only be answered by increasing the number of ETGs observed with spatially-resolved spectroscopy and comparing these observations to a variety of simulations such as those shown in \citet{Bois2011}, \citet{Khochfar2011} and \citet{Naab2013}. Since the collection of integral field spectroscopy (IFS) data tends to be time-consuming these observations can be difficult to accomplish quickly. However, the introduction of new multi-object IFS instruments means that soon large samples of galaxies will be available for study in this way.

The Sydney-AAO Multi-object Integral field spectrograph \citep[SAMI;][]{Croom2012} is one such multi-object IFS. SAMI was commissioned at the Anglo-Australian Telescope (AAT) in 2011 and comprises 13 fibre bundle integral field units (IFUs) deployable across a one-degree diameter field of view. SAMI can target 13 galaxies in a single observation (or more usually 12 galaxies and one calibration star), significantly decreasing the amount of time needed to build a large sample of galaxies with IFS data. A large-scale galaxy survey using SAMI is currently underway (the SAMI Galaxy Survey\footnote{http://sami-survey.org}) with many scientific aims surrounding the nature of galaxy formation and evolution. 

The SAMI Pilot survey is a precursor to the SAMI Galaxy survey, comprising observations of three galaxy clusters, Abell 85, Abell 168 and Abell 2399. The SAMI Pilot survey was carried out in order to answer some of the specific questions about ETGs in clusters mentioned above. This paper focusses on one aspect of this, investigating the kinematic morphology-density relation in the three clusters. This work almost doubles the number of clusters previously examined in this way.

Throughout we adopt a cosmology with $\Omega_{m}=0.3$, $\Omega_{\Lambda}=0.7$ and H$_{0}=70\ \rm{km\ s^{-1}\ Mpc^{-1}}$.

\section{Observations and Data}
\label{sec:obs}

We use data from SAMI on the AAT, along with archival data from the Sloan Digital Sky Survey (SDSS) and the {\it Chandra} and {\it XMM Newton} telescopes.

\subsection{SAMI Data}

SAMI \citep{Croom2012} is a multi-object integral field spectrograph mounted at the prime focus of the AAT. SAMI's 13 IFUs (called hexabundles, \citet{JBHBryant2011, Bryant2011, Bryant2014}) are deployable across the 1-degree diameter field of view provided by the AAT triplet corrector. Each hexabundle subtends $\sim15\arcsec$ in diameter on the sky, and comprises 61 individual fibres each with a core diameter of 1.6$\arcsec$. The circular fibres are arranged in a circular pattern, so inevitably the spatial coverage of a hexabundle is not contiguous. However, the SAMI hexabundles have a high fill factor, with a mean of 73\%. As well as 13 hexabundles, SAMI also has 26 individual sky fibres, deployable across the field of view. This enables accurate sky subtraction for all IFU observations without the need to observe separate blank sky frames. Since an individual SAMI fibre undersamples the seeing at Siding Spring ($1.9\arcsec-3.0\arcsec$ during our observations) we adopt a dithering strategy to regain spatial resolution in our final combined data cubes. 


From the AAT prime focus unit, SAMI's 819 fibres (including both the hexabundle fibres and the dedicated sky fibres) feed the double-beamed AAOmega spectrograph \citep{Sharp2006}.  AAOmega is a fully configurable spectrograph with multiple dichroics and diffraction gratings to choose from. The standard configuration for SAMI uses the 580V grating in the blue arm, giving R $\sim$ 1700 over the wavelength range 3700\AA\ - 5700\AA. This configuration was chosen to cover a broad range of important stellar absorption features, including the \caii\, H+K,  \hb\, and Mgb lines in galaxies up to a redshift of 0.1. In the red arm SAMI uses the 1000R grating, giving R $\sim$ 4500 over the wavelength range 6250\AA\ - 7350\AA. This configuration was chosen to allow high-resolution observations of the \ha\, and \nii\, emission line complex. For this work we are primarily interested in fitting the stellar absorption features covered by the blue arm of AAOmega. We therefore analyse only the blue arm SAMI data.

\subsubsection{The SAMI Pilot Survey ETGs}
\label{sec:pilotsurvey}
The observations presented and analysed in this work form the SAMI Pilot Survey, a smaller, more targeted precursor to the full SAMI Galaxy Survey. The SAMI Pilot Survey was focussed on observations of galaxies in cluster environments.

A parent sample of clusters was identified from the catalogue of \citet{Wang2011arXiv}, constrained to have $z<0.06$ and to be observable at the AAT (i.e. $\rm{Dec.}<10^{\circ}$). Galaxies were then included in the sample if they were within 1$^{\circ}$ radius of the cluster centre and a redshift range of $0.025<z<0.085$ and had M$_r<-20.25$. This selection includes foreground and background objects in the galaxy sample for each cluster. Seven clusters were included in the parent sample and three of these were observed: Abell 85, Abell 168 and Abell 2399. Table \ref{tab:clusters} gives some details about each of the three clusters. 

For the SAMI Pilot survey a total of 14 galaxy fields were observed on 10 nights across two separate runs in September and October 2012. Due to hardware problems which have since been solved, we used only 9 hexabundles per field, observing 8 galaxies and one calibration star in each. We observed a total of 112 galaxies. However, data for 6 of the galaxies proved to be unusable due to astrometric errors during the September 2012\, run. The final observed sample contains 106 galaxies, of all morphological types, in three cluster fields. 

For this work we are primarily interested in the ETGs so before performing our analysis we morphologically classified each galaxy by eye. The purpose of this step is not to generate detailed classifications but only to reject spiral and irregular galaxies from our sample. For that reason, our classification system contains only two categories, ETGs and LTGs. We define an ETG to be a galaxy which has a smooth and symmetric light profile and does not exhibit spiral structure. We attempt to match the ATLAS$^{\rm{3D}}$ classification criteria as much as possible, but this is inevitably imperfect since our sample is much more distant. We used SDSS DR8 $r$-band images, consulting SDSS colour ($gri$) images in borderline cases. All galaxies with obvious spiral arms were excluded but if a galaxy was borderline it was included in the sample. This yielded 79 ETGs and 27 LTGs in our sample of 106 galaxies. We focus only on the ETGs in this paper, and will deal with the LTGs in Fogarty et al. in prep. (hereafter Paper I).

The final sample of ETGs includes 5 galaxies which are not cluster members, but foreground or background interlopers. The criteria by which we allocate cluster members is discussed in Section \ref{sec:clustermembership}. The 5 non-members in our sample are included in our analysis, but left out of any cluster-related statistics or measurements. This leaves us with a sample of 74 cluster member ETGs. The distribution of the ETGs by cluster is summarised in Table \ref{tab:clusters}.

\begin{table*}
\centering
\begin{tabular}{|c|c|c|c|c|c|c|c}
\hline
Cluster & R.A. (J2000) & Dec. (J2000) & Redshift & M$_{200}$ Virial & L$_{\rm{X}, 500}$ & Number of  & Number of  \\
 & & & & (x\,10$^{14}$\,M$_{\odot}$) & (x\,10$^{44}$ erg\,s$^{-1}$) & ETGs & Cluster ETGs \\
\hline
Abell 85 & 10.46021 & -9.30318 & 0.0549 & $11.9\pm1.4$ & 5.10 & 28 & 28 \\
Abell 168 & 18.73997 & 0.40807 & 0.0451 & $2.4\pm0.3$ & 0.47 & 12 & 12 \\
Abell 2399 & 329.38949 & -7.79424 & 0.0583 & $5.3\pm0.7$ & 0.45 & 39 & 33 \\
\hline
All & -- & -- & -- & -- & -- & 79 & 73 \\
\hline
\end{tabular}
\caption{Information about the three clusters observed as part of the SAMI Pilot Survey. Note that the three clusters have a range in mass and X-ray luminosity (columns 5 and 6). The X-ray luminosities are from \citet{Piffaretti2011} Roughly half the observed sample is part of the Abell 2399 selection and this is the only cluster sample with contamination from non-member ETGs.}
\label{tab:clusters}
\end{table*}

\subsubsection{SAMI Data Reduction}
\label{sec:samidr}

The data reduction procedure for the SAMI Galaxy Survey is described in detail in two forthcoming papers Allen et al. in prep. and Sharp et al. in prep.. The data from the SAMI Pilot Survey are similar but not identical to the SAMI Galaxy Survey data. We follow the same data reduction procedure for both campaigns except when performing telluric correction (which is not relevant for this work, which is confined to the blue SAMI data cubes) and with one minor difference when combining dithers to create the final data cubes. The procedure is described in detail the two papers mentioned above and is therefore described only briefly below.

All SAMI data are reduced from raw files to row-stacked spectrum (RSS) format using the \twodfdr\, data reduction package \citep{SharpBirchall2010}. \twodfdr\, was initially written to process data from the 2dF spectrograph and has been updated to process all AAOmega data, including SAMI data. \twodfdr\, performs bias subtraction and pixel-to-pixel flat-fielding before extracting individual fibre spectra from a raw frame. The extracted spectra are then wavelength calibrated, sky subtracted and corrected for fibre-to-fibre throughput variations, before being reformatted into a RSS file, consisting of all 819 extracted SAMI spectra. Variance information is also propagated by \twodfdr\, so output fibre spectra each have corresponding variance spectra. After the RSS files are produced an observation of a spectrophotometric standard star is used for flux calibration. A separate calibration star, observed at the same time as the galaxy field is used to correct for telluric absorption. For the SAMI Pilot Survey the calibration star had low quality data and so these data were treated differently to the data in the SAMI Galaxy Survey. However, this is not relevant to the results presented in this paper as the telluric correction affects the red arm data cubes only and so a description is omitted here.

Since the SAMI hexabundles have non-uniform spatial coverage, all galaxy fields are observed multiple times with a small dither between exposures. The purpose of this observing strategy is to fill in the gaps between fibres and regain spatial resolution. For the SAMI Galaxy Survey a fixed seven-point hexagonal dither pattern with an offset of 60\% of a core diameter between dithers is used (see Sharp et al. in prep. for details). However during the SAMI Pilot Survey the dither pattern was not set and the step between dithers was larger. This impacts the chosen ``drizzle'' parameters applied to this data set which do not match those chosen for the SAMI Galaxy Survey.

Once the flux-calibrated RSS frames (containing both data and variance spectra) for each dither are in hand, these are combined to form data and variance cubes for each galaxy in the observed field. This step is performed using custom python code developed by the SAMI Galaxy Survey team and the reader is referred to Sharp et al. in prep. for a detailed discussion of how this is acheived. In short, the round SAMI fibres (1.6$\arcsec$ diameter) are resampled onto a regular grid of square output spaxels using an algorithm similar to Drizzle \citep{FruchterHook2002}. The fibres are shrunk by some user-chosen drop factor and ``drizzled'' onto the regular output grid. For the SAMI Pilot Survey the drop factor is 0.75 and the output spaxels are 0.5$\arcsec$ on each side. All dither positions are resampled in this way and combined at each wavelength slice, with a correction for differential atmospheric refraction. The variance spectra are also propagated through this process. This process yields two data cubes for each galaxy, one each for the blue and red spectrograph arms, along with two corresponding variance cubes. 

It is important to note that this resampling scheme generates data and variance cubes in which individual spaxels are not independent. The individual variance spectra give the correct uncertainties for individual data spectra, but since light from a single observed spectrum may contribute to more than one output spectrum, the uncertainties in adjacent output spaxels are correlated. This has implications when measuring uncertainties on integrated properties (such as $\lambda_{R}$) or when calculating the true variance on composite spectra formed by summing two or more different spaxels. In a single wavelength slice, each spaxel is correlated with some number of its nearest neighbours. The strength of this correlation depends on the position of the original fibres in relation to the output spaxel grid. The correlation information for each spaxel is recorded for a 5 x 5 grid of its surrounding spaxels only. This information is carried for each spaxel in each wavelength slice, yielding a third set of  ``cubes'' (these have two extra dimensions covering the 5 x 5 grid) of correlation information for each galaxy. These cubes must be multiplied by the variance in the spaxel(s) of interest to reconstruct the true variance. For example, consider the case where we wish to sum the spectra in two neighbouring spaxels and estimate the true uncertainly in the resulting spectrum. We must first isolate the spaxels of interest in both relevant 5 x 5 correlation factor grids, multiply by the appropriate variance spectra and then sum the result in order to reconstruct the complete, true variance spectrum (including the covariance between contributing spaxels) appropriate to the new composite spectrum.

The method we use to account for this effect when we calculate the uncertainty in $\lambda_{R}$ is detailed in Section \ref{sec:errors}.

\subsection{Archival Data}

\subsubsection{X-Ray Imaging}

\begin{figure*}
\centering
\includegraphics[width=10cm]{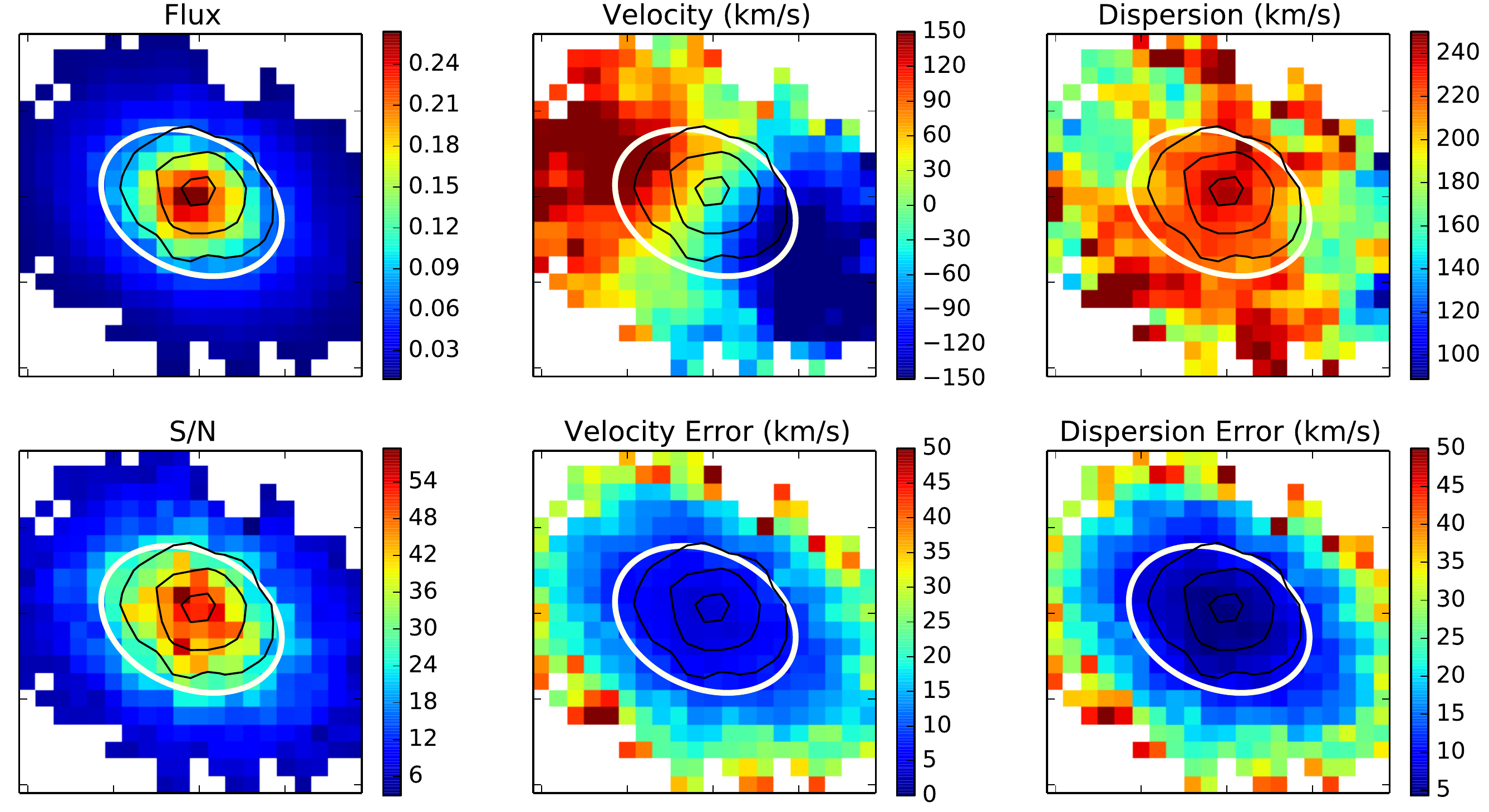}
\caption{This figure shows the stellar kinematic maps for J004001.68-095252.5, a member of Abell~85. The maps were derived using pPXF \citep{CappellariEmsellem2004} using the 985 MILES stellar templates \citep{SanchezBlazquez2006} as reference spectra. The top row shows, from left to right, summed flux in arbitrary units, V in km/s and $\sigma$ in km/s. The bottom row shows, left to right, S/N measured across 200 wavelength slices of the data cube, uncertainty on V in km/s and uncertainty in $\sigma$ in km/s. The black contours indicate measured flux from the SAMI cube, i.e. that shown in the top-left panel. Each panel is 10$\arcsec$ on a side. The ivory coloured ellipse represents one effective radius.}
\label{fig:kin_example}
\end{figure*}

Both Abell~85 and Abell~168 have extensive X-ray data available in the {\it Chandra} archives. The data for Abell~85 consists of nine pointings using the ACIS-I chip array: one $\sim 38\,$ks exposure centred on the BCG \citep[ObsID 904;][]{Kempner2002} and eight $\sim 10\,$ks exposures in the peripheral cluster regions \citep[ObsIDs 4881, 4882, 4883, 4884, 4885, 4886, 4887, 4888;][]{Sivakoff2008}. For Abell~168, there are two ACIS-I {\it Chandra} observations with $\sim 37\,$ks and $\sim 40\,$ks exposures \citep[ObsIDs 3203 and 3204;][]{Hallman2004}. The {\it Chandra} data were reprocessed using the CHANDRA\_REPRO script within the CIAO software package \citep[version 4.4;][]{Fruscione2006}. The script applies the latest calibrations to the data (CALDB 4.5.1), creates an observation-specific bad pixel file by identifying hot pixels and events associated with cosmic rays (utilizing VFAINT observation mode where available), and filters the event list to include only events with {\it ASCA} grades 0, 2, 3, 4, and 6. The DEFLARE script is then used to identify periods contaminated by background flares. No significant contamination was found in either data set. For the imaging analyses, exposure maps which account for the effects of vignetting, quantum efficiency (QE), QE non-uniformity, bad pixels, dithering, and effective area were produced using standard CIAO procedures\footnote{cxc.harvard.edu/ciao/threads/expmap\_acis\_multi/}. The energy dependence of the effective area is accounted for by computing a weighted instrument map with the SHERPA {\sf make\_instmap\_weights} script using an absorbed MEKAL spectral model with temperature, abundance and neutral hydrogen density appropriate for each cluster. Background subtraction was performed using the blank sky backgrounds\footnote{cxc.harvard.edu/contrib/maxim/acisbg/}$^,$\footnote{cxc.harvard.edu/caldb/downloads/Release\_notes/supporting/\\README\_ACIS\_BKGRND\_GROUPF.txt} which were processed in the same manner as the observations. The blank sky backgrounds were reprojected to match the tangent point of the observations, and were normalized to match the $9-12\,$keV counts in the observations. This yielded clean, calibrated X-ray images for Abell~85 and Abell~168.

Abell~2399 has three archival {\it XMM-Newton} observations (ObsIDs 0201902801, 0404910701 and 0654440101). Here, we use only the June 2010 (0654440101) observation as it has the longest exposure ($\sim 93\rm{ks}$). The raw  data are processed following the {\it XMM-Newton} Extended Source Analysis Software (XMM-ESAS) package\footnote{ftp://xmm.esac.esa.int/pub/xmm-esas/xmm-esas.pdf} \citep[see also][]{Snowden2008} within the Science Analysis System (SAS)\footnote{http://xmm.esac.esa.int/sas/}. The tasks {\it emchain} and {\it epchain} are run on the MOS and PN data to produce event lists filtered of bad pixels and with the latest calibrations applied. The {\it pn-filter} and {\it mos-filter} tasks are used to identify and remove periods of increased background due to flare activity. The data were moderately affected by flares and the cleaned exposure times are 54\,ks, 60\,ks and 26ks for the MOS1, MOS2 and PN detectors, respectively. Images and exposure maps were generated by the {\it mos-spectra} and {\it pn-spectra} tasks using the cleaned event lists in the $0.4-1.25\,$keV and $2.0-7.2\,$keV bands (to avoid the strong instrumental lines). Corresponding maps of the quiescent particle background were produced with the {\it mos-back} and {\it pn-back} tasks. These background maps are recast to the source image coordinates using the {\it rot-img-det-sky} task. Source, background and exposure images for all bands and detectors were combined with the {\it comb} task, yielding a final X-ray image of Abell~2399. The resulting X-ray images are discussed in Section \ref{sec:results}.

\subsubsection{SDSS Optical Imaging}

As discussed in Section \ref{sec:pilotsurvey} we use SDSS DR8 $r$-band postage stamps and $gri$ colour images of each galaxy to morphologically classify our sample. The $r$-band postage stamps are also used to derive photometric parameters, such as effective radius, for each object. The procedure followed is described in Section \ref{sec:photometry}.

In addition, three-colour mosaics were created from SDSS DR7 $gri$ images of each cluster using Montage. The mosaics cover the central parts of each cluster and are presented in Section \ref{sec:results}.

\section{Derived Parameters}
\label{sec:derived}

\subsection{Stellar Kinematics}
\label{sec:ppxf}

We fit stellar kinematic fields for each of our 79 ETGs using the penalised pixel-fitting routine, pPXF, created by \citet{CappellariEmsellem2004}. pPXF uses a penalised maximum likelihood method to fit stellar template spectra convolved with an appropriate line-of-sight velocity distribution (LOSVD) to observed galaxy spectra. The LOSVD is parametrised by a truncated Gauss-Hermite expansion, allowing higher orders of the LOSVD to be fit, beyond velocity (V) and velocity dispersion ($\sigma$). Here we fit four LOSVD moments: V, $\sigma$, h$_3$ and h$_4$, though in practice we use only the first two of these for our analysis. We use the 985 MILES \citep{SanchezBlazquez2006} stellar templates as reference spectra.

For each of our galaxies the procedure is as follows. First a high signal-to-noise (S/N) spectrum is extracted from the blue data cube in a 2$\arcsec$ circular aperture centered on the galaxy. The pPXF routine is run on this spectrum to find the best fit templates for that galaxy. This reduced set of templates are then fit to individual spaxels within the data cube, with the weights given to each template allowed to vary. An additional fourth order polynomial was fit with the templates in order to account for any residual flux calibration errors in our SAMI data. Only spaxels for which the S/N was greater than 5 per spectral pixel were fit. This produces maps of V, $\sigma$, h$_3$ and h$_4$ for each galaxy, along with maps of the uncertainties on these quantities. The uncertainties calculated from the pPXF fits are correct for individual spaxels but are correlated with adjacent spaxels, an effect which impacts any integrated parameters derived from the kinematic maps. 

An example set of kinematic maps for J004001.68-095252.5, a member of Abell 85, are shown in Figure \ref{fig:kin_example}. The top three panels show the flux, V and $\sigma$ maps generated by pPXF. The bottom left panel shows a S/N map and the bottom centre and right panels show the uncertainties on V and $\sigma$ respectively (each panel is 10$\arcsec$ on a side). Kinematic maps for all galaxies in the sample, including the 27 LTGs, will be presented in Paper I.

\subsection{Photometric Fitting}
\label{sec:photometry}

We derive several photometric parameters for each galaxy in our sample using the SDSS DR8 $r$-band images. For the analysis presented here we are particularly interested in accurate values for the effective radius, R$_e$, and the ellipticity ($\epsilon$) and position angle (PA) for each galaxy.

Using the approach described in \citet{Cappellari2007} we construct a Multi Gaussian Expansion (MGE) model \citep{Emsellem1994} of the surface brightness profile from the SDSS image of each galaxy. The surface brightness profile is described as the sum of a set of two dimensional Gaussians with varying normalisation, width and axial ratio. This formalism is extremely flexible and provides a good description of the surface brightness profiles of a wide variety of galaxy morphologies \citep{Scott2013}. From the MGE model we determine the effective radius following the method of \citet{Cappellari2009}. The MGE model is circularised, setting the axial ratios of each Gaussian component to 1 while preserving the peak surface brightness of each component, then the radius which encloses exactly half the total luminosity of the model is determined by interpolating over a grid of radial values.

We determine ellipticity and position angle profiles directly from the SDSS images using the {\sc idl} routine {\it find\_galaxy.pro} \citep{Krajnovic2011}. The ellipticity and position angle at a given radius are determined from the second moments of the luminosity distribution of all connected pixels above a given flux level. For example, $\epsilon_\mathrm{e}$ and PA$_\mathrm{e}$ are the values of the ellipticity and position angle determined at the isophote that encloses an area equal to $\pi$R$_e^2$. 

\subsection{Cluster Membership}
\label{sec:clustermembership}

\begin{figure}
\centering
\includegraphics[width=8.5cm]{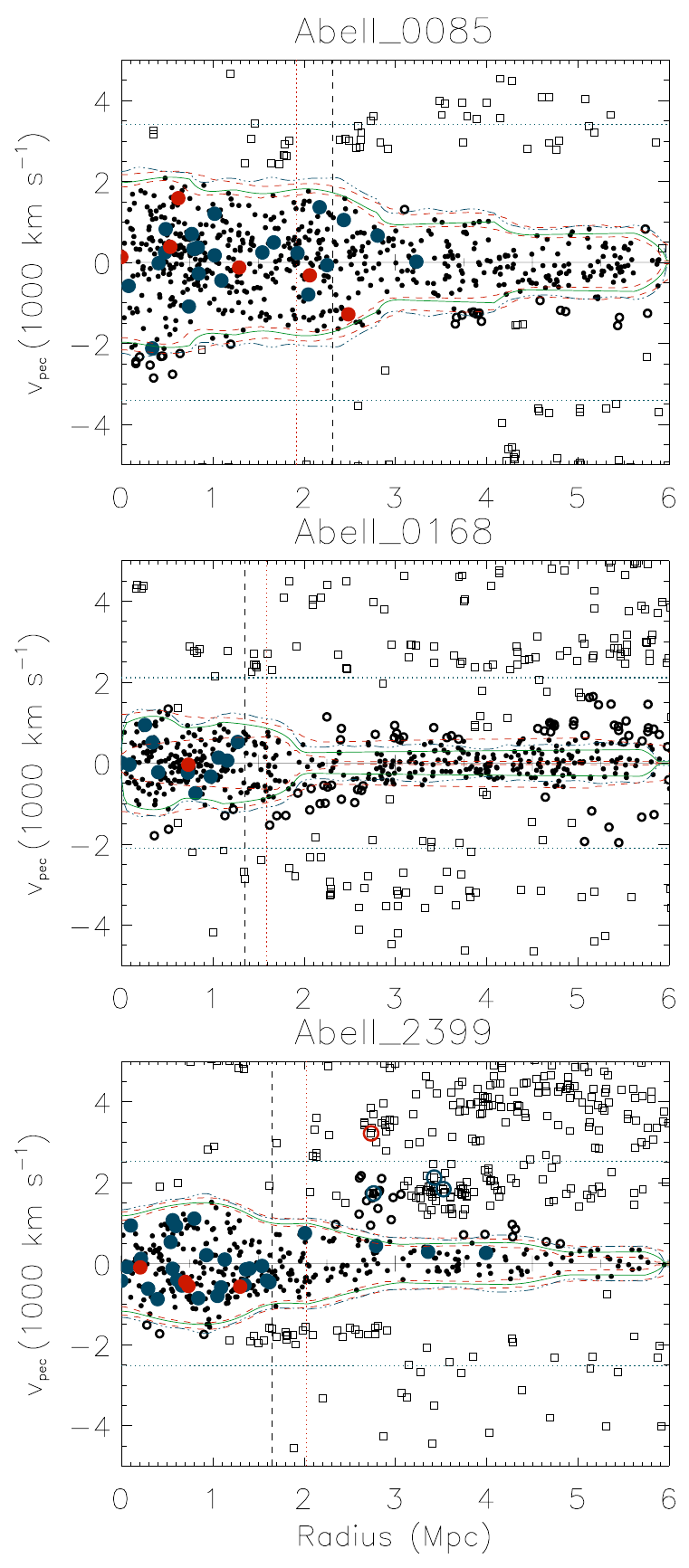}
\caption{This figure shows the phase space distribution of galaxies in each of the
three clusters in the SAMI Pilot Survey. Filled circles show the
spectroscopically confirmed cluster members. The black open squares show
the non-members identified by the shifting-gapper method. The black open
circles show the non-members identified as being beyond the limits defined
by the velocity caustics (green solid line). The red dashed line shows the
68th percentile uncertainties on the caustics. The blue dot-dashed
line shows the standard analytic estimate of the uncertainty in the
caustic measurement. The larger blue and red circular points (both open
and closed) are the ETGs observed in the SAMI Pilot Survey, with colours
indicating their kinematic classification (see Section 4).  The vertical
dashed black lines show R200 and the vertical red dotted lines show the SAMI field of
view.}
\label{fig:phase_space}
\end{figure}

To study the distribution of kinematic morphologies for the three clusters in our sample, we must ensure we have an accurate determination of cluster membership for all 79 ETGs. If we do not reliably determine which galaxies are cluster members we will bias our analysis by including line of sight interlopers. The three clusters from the SAMI Pilot Survey are included in the SAMI Cluster Redshift Survey (Owers et al. in prep.). The survey was carried out using the 2dF/AAOmega multi-object fibre spectrograph, capable of targeting $\sim392$ galaxies in one exposure. The survey provides deeper spectroscopy and therefore greater completeness than that available from SDSS. Therefore, we use the catalogue derived from the SAMI Cluster Redshift Survey to determine cluster membership for our sample.

The allocation of cluster members is an iterative process which will be
described in detail in a forthcoming paper (Owers et al. in prep.).
Briefly, an initial cull of non-members is achieved by rejecting those galaxies
which have $|v_{\rm pec}| < 5000\,$km/s where $v_{\rm pec}$ is the
peculiar velocity measured with respect to the redshift of the BCG.
Following this initial rejection, the process enters an iterative phase
where a modified version of the  ``shifting-gapper''  method
\citep{Fadda1996, Owers2009} is used to refine the membership selection.
The galaxies are sorted by projected distance from the BCG and split into
annular radial bins. The width of the annular bins is set such that each
bin contains 40-50 galaxies. Within each radial bin, the galaxies are
sorted by $v_{\rm pec}$, and the velocity gap, $\Delta v_{\rm pec}$,
between consecutive galaxies is determined. The distribution of $\Delta
v_{\rm pec}$ is searched for values which exceed a threshold value,
$\Delta v_{\rm thresh}$, which is taken to be the velocity dispersion
within the annular bin of interest. This local velocity dispersion is
estimated from the median absolute deviation which is a robust,
outlier-resistant measure of the scale of a distribution
\citep{Beers1990}. Galaxies which have gap values exceeding $\Delta v_{\rm
thresh}$, as well as having $|v_{\rm pec}| > \sigma_{\rm v}$, where
$\sigma_{\rm v}$ is the cluster velocity dispersion, mark the outer limits
of the cluster in the annular radial bin of interest. Galaxies with larger
velocities are removed as interlopers. The procedure is iterated until the
number of members stabilises. Galaxies identified as non-members using
this method are shown in Figure \ref{fig:phase_space} as black open squares.

In a few cases, structure which is nearby in redshift, but is unlikely to
be within the cluster, causes the shifting-gapper method to fail (e.g., in
Figure \ref{fig:phase_space},  $\sim 2.8\,$Mpc from the BCG in A2399). To address this, we use
the adaptively-smoothed distribution of galaxies in $v_{\rm pec}$-radius
space to locate the cluster caustics \citep{Diaferio1999}. The caustics,
shown as green solid lines in Figure~2, trace the escape velocity of the
cluster as a function of cluster-centric radius and, therefore, robustly
identify the boundary in $v_{\rm pec}$
radius space between bona-fide cluster members and line-of-sight
interlopers \citep[e.g.,][]{Serra2013, Owers2013}. Two measures of the
uncertainty on the caustic boundary are measured. The first, shown as a
blue dash-dot line in Figure \ref{fig:phase_space}, is the analytical approximation to the
$1-\sigma$ uncertainty and is measured as described by
\citet{Diaferio1999}. The second, shown as red dashed lines in Figure \ref{fig:phase_space}
shows the 68th percentile limits derived from a distribution of caustic
measurements determined from 1000 boostrap resamplings of the cluster
member phase-space distribution. Only galaxies which lie within the outer
68th percentile limit are retained in the final cluster member sample.
Those galaxies rejected using this method are shown as black open circles
in Figure \ref{fig:phase_space}. Using the sample of spectroscopically confirmed cluster
members within $R_{200}\simeq
\sqrt(3) \sigma_{\rm v}/10H(z)$, the virial masses listed in Table \ref{tab:clusters} were
determined using the method outlined in \citet{Owers2009}.

\subsection{Galaxy Environment}

In any discussion of the influence or importance of galaxy environment one must first think carefully about the meaning and quantification of ``environment''. Here we consider two types of ``environment''. Following \citet{Houghton2013} we define the ``global host environment'' (GHE) to refer to the global environment of a galaxy - i.e. does it reside in a cluster, group, or in the field? Another consideration here is the overall mass and density of individual clusters when compared to one another.

We also use the ``local point environment'' (LPE) as per \citet{Houghton2013} to mean the local projected surface density at the position of a galaxy. In this paper we will focus more on this measure of environment, since the GHE for all of our galaxies is similar - they all reside in clusters, albeit ones with differing overall cluster properties.

These two definitions are helpful when considering the overall position of individual galaxies. As an example, do galaxies with the same measured projected environment (i.e. LPE) behave the same regardless of whether they are central in a group GHE or on the outskirts of a cluster GHE? This question is beyond the scope of this paper but it concisely illustrates the need to think of ``environment'' in several complex and possibly interdependent ways.

\subsubsection{Local Point Environment Measurements}
\label{sec:lpe}

We derive a value for the nearest neighbour surface densities ($\Sigma_N$) for all galaxies in the SDSS Stripe 82 observations that cover these clusters. The surface density is defined using the projected co-moving distance to the Nth nearest neighbour ($d_{N}$) with velocity=$\pm2000 \rm{\ km\ s}^{-1}$ within a volume limited density-defining population: $\Sigma_N=N/\pi d^2_N$. The density-defining population has absolute SDSS petrosian magnitudes M$_r<$M$_{r,\rm{limit}}$-Q$z$, k-corrected to z = 0, where  M$_{r,\rm{limit}}$= -19.0 mag and Q defines the expected evolution of $M_{r}$ as a function of redshift (Q=0.87; \citet{Loveday2012}).  We then calculate densities for N=3, N=5 and N=10. To aid comparison with ATLAS$^{\rm{3D}}$ we have used the same N, velocity and absolute magnitude limits. 

\section{Kinematic Classification}
\label{sec:rotators}

\begin{figure*}
\centering
\includegraphics[width=17cm]{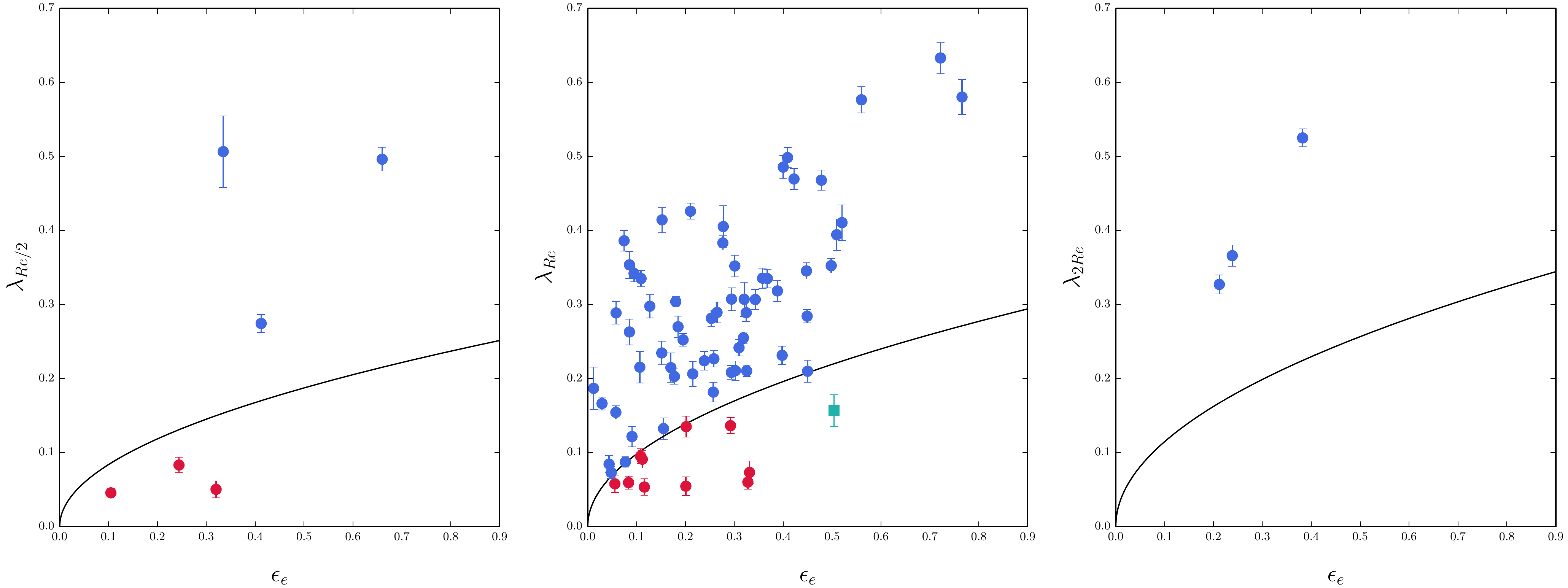}
\caption{This plot shows $\lambda_{R}$ plotted against $\epsilon$ for all 79 ETGs in the SAMI Pilot sample. The left-hand panel shows those galaxies classified at R$_{e}/2$, the middle shows those classified at R$_{e}$ and the right-hand panel shows those classified at 2R$_{e}$. In all panels the black line shows the division between FRs and SRs as defined by Equation \ref{eqn:line}. The blue points are FRs and the red are SRs. The square turquoise coloured point in the centre panel is J215634.45-075217.5. This is a double sigma galaxy (not a true SR) and is discussed in Section \ref{sec:other_class}.}
\label{fig:lambda_multiline}
\end{figure*}

We wish to classify our sample of 79 ETGs as slow or fast rotators (SR/FR). In this section we examine several methods of kinematic classification, both quantitative and qualitative. 

\subsection{Aperture Effects}

To quantitatively classify our galaxies we first measure $\lambda_{R}$ from our derived kinematic maps. $\lambda_{R}$ in a proxy for the luminosity-weighted specific stellar angular momentum for each galaxy within a fiducial radius and is defined as follows \citep{Emsellem2007}:

\begin{equation}
\lambda_R\equiv\frac{\langle R|V|\rangle}{\langle R\sqrt{V^2+\sigma^2}\rangle}=\frac{\Sigma_{i=0}^{N}F_i R_i |V_i|}{\Sigma_{i=0}^{N}F_i R_i \sqrt{V_i^2+\sigma_i^2}}
\end{equation}

\noindent where $F_i, R_i, V_i, \sigma_i$ are the flux, radius, velocity and velocity dispersion of the $i$th of $N$ spaxels included in the sum. 

Ideally we would like to measure $\lambda_{R}$ within the effective radius, R$_{e}$, for all 79 ETGs in our sample. Although our sample covers a range in angular size, the majority of galaxies have $2\arcsec\leq \rm{R}_e\leq7\arcsec$ and fit comfortably within the SAMI hexabundles with sufficient spatial resolution. For these 70 objects we calculate $\lambda_{R}$ within R$_e$.

The six largest galaxies in the sample have R$_e>7\arcsec$ and overfill the SAMI hexabundles. For these objects we measure $\lambda_{R}$ at $R_{e}/2$ instead. \citet{Emsellem2007} showed, using the SAURON Survey sample, that it is rare for a galaxy to change its classification based on $\lambda_R$ between R$_e/2$ and R$_e$.

Conversely, three galaxies in our sample are quite small, such that R$_e$ covers only one or two independent resolution elements. In this case a measurement of $\lambda_{R}$ within R$_e$ will be biased to low values. This is because a velocity gradient across a single resolution element will be unresolved and measured as dispersion. To avoid this effect we measure $\lambda_{R}$ within 2R$_e$ for the three galaxies with R$_e<2\arcsec$. Unlike measurements of $\lambda_{R}$ within R$_e/2$ and R$_e$, the properties of this parameter have not been fully investigated at large radii. However, the compact galaxies in our sample display very orderly velocity maps and good S/N out to large radii. They are all very clear FRs with no ambiguity in their classifications, so we feel justified in extending the $\lambda_{R}$ measurements of these objects to 2R$_e$.

Despite the fact that we use three fiducial radii for our sample, Figure \ref{fig:lambda_multiline} clearly demonstrates that for our purposes this does not influence our classifications. The galaxies with $\lambda_{R}$ measurements within a smaller (R$_e/2$) and larger (2R$_e$) radii than originally used to define the classification system are all unambiguously positioned in $\lambda_{R}-\epsilon$ space. That is to say they fall away from the dividing line and there is no doubt that the system adopted here works well for this sample.

\subsection{Classification Using $\lambda_{R}$}
\label{sec:lambdar}

The measured flux, V and $\sigma$ maps are used to calculate $\lambda_{R}$ parameter at the appropriate fiducial radius for each galaxy. The calculation uses the R$_e$, $\epsilon$ and PA measurements from the MGE fits described in Section \ref{sec:photometry}. The $\epsilon$ and PA values used are those measured at the correct fiducial radius for each galaxy. The results are presented in Table \ref{tab:all_gals}.

To discriminate between FRs and SRs we use the $\lambda_{R}-\epsilon$ space. Since we have used three fiducial radii to measure $\lambda_{R}$ for our sample, we need three different criteria to divide FRs and SRs. We use the criterion defined by \citet{Emsellem2011}, which is dependent on ellipticity:

\begin{equation}
\lambda_{R}<k\sqrt{\epsilon}
\label{eqn:line}
\end{equation}

\noindent where $\lambda_{R}$ and $\epsilon$ are measured within the same fiducial radius in the galaxy and the value of $k$ is dependent on the choice of fiducial radius for a particular galaxy. Assuming a linear scaling from the values of $k$ at R$_{e}/2$ and R$_{e}$ published in \citet{Emsellem2011}, we adopt $k=0.265$, $k=0.310$ and $k=0.363$ for values of $k$ at R$_{e}/2$, R$_{e}$ and 2R$_{e}$ respectively.

Figure \ref{fig:lambda_multiline} shows $\lambda_{R}-\epsilon$ for all 79 ETGs in the SAMI Pilot Survey. The three panels correspond to the three fiducial radii used to measure $\lambda_{R}$ and $\epsilon$. The left-hand panel shows the galaxies classified at R$_{e}/2$, the centre panel shows those classified at R$_{e}$ and the right-hand panel shows those classified at 2R$_{e}$. In all three panels the black solid line marks the division between FRs and SRs, such that SRs lie below the line and FRs above it. Thus the blue points are classified as FRs and the red as SRs.

\begin{figure}
\centering
\includegraphics[width=8.5cm]{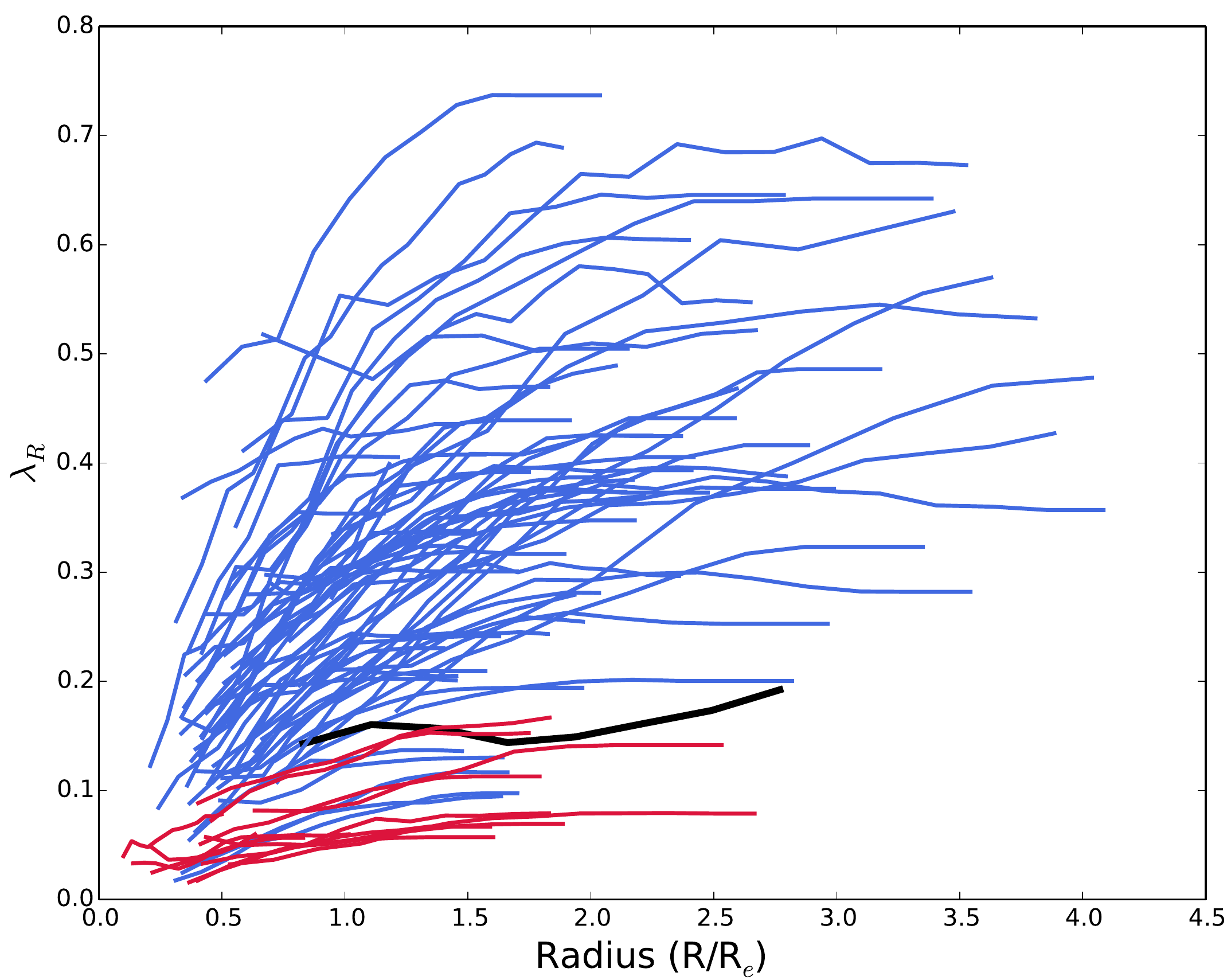}
\caption{The radial profiles of $\lambda_{R}$ for all 79 ETGs in our sample. The blue lines indicate FRs and the red lines indicate SRs. The black line highlights J215634.45-075217.5, the double sigma galaxy discussed in Section \ref{sec:other_class}. The x-axis shows galaxy radius in units of effective radius, R$_e$, clearly illustrating the differing spatial coverage of our data.}
\label{fig:radial_lr}
\end{figure}

We also calculate resolved radial profiles of $\lambda_{R}$ for each galaxy. These are shown in Figure \ref{fig:radial_lr} where FRs are shown in blue and SRs in red, as classified using the $\lambda_{R}-\epsilon$ space. Of course, since $\lambda_{R}$ is a projected parameter, and the radial profile does not include any information about the ellipticity of the galaxies, these profiles do not give the whole picture. However, the overall shapes of profiles can be helpful as a check on other methods of classification. We expect that for FRs $\lambda_{R}$ will increase rapidly with radius, perhaps with a plateau at large radii. For SRs we expect a decrease or a very slow increase in $\lambda_{R}$. The latter could be caused by increased noise at higher radii which can bias measurements of $\lambda_{R}$ to larger values. In our sample a range of profile shapes can be seen, with no clear gap between the two populations of objects. The SRs tend have shallower slopes than the FRs but there is some clear overlap in the two classes of objects. The overlap is caused by the fact that $\lambda_{R}$ is a projected quantity and therefore information about the ellipticity of individual galaxies is needed to fully classify them as SRs or FRs. Galaxies in this overlap region may have similar values for $\lambda_{R}$ at a particular radius but differences in ellipticity will mean they can have different kinematic classifications. This illustrates the need to consider more than one method of kinematic classification of ETGs.

\subsection{Visual Classifications}
\label{sec:other_class}

We also classify galaxies as FRs and SRs based on visual inspection of the kinematic fields. Several authors examined the kinematic maps for all 79 ETGs in the sample. This was done in isolation and without prior knowledge of the $\lambda_{R}-\epsilon$ classifications. The kinematic maps inspected were identical to those in the top row of Figure \ref{fig:kin_example}. The error maps were not provided. The visual classifications were carried out using a scheme of five classes designed to incorporate some uncertainty. The classifications were, definite FR, probable FR, unknown, probable SR, definite SR. The results were collated and compared after all authors completed their individual classifications. For most galaxies all or a majority of authors agree on the visual classification. In most of these cases the visual and $\lambda_{R}-\epsilon$ classifications also agree. 

In cases where the visual and $\lambda_{R}-\epsilon$ classifications do not agree the galaxy in question is usually close to the dividing line in $\lambda_{R}-\epsilon$ space. For our statistical analysis we adopt the $\lambda_{R}-\epsilon$ classification in all such cases, of which there are only a handful. However, there are some galaxies of interest for which it is useful to discuss both the quantitative $\lambda_{R}-\epsilon$ classifications and the contradicting visual classification. One such galaxy is discussed in Section \ref{sec:a168}.

\begin{figure}
\centering
\includegraphics[width=8.5cm]{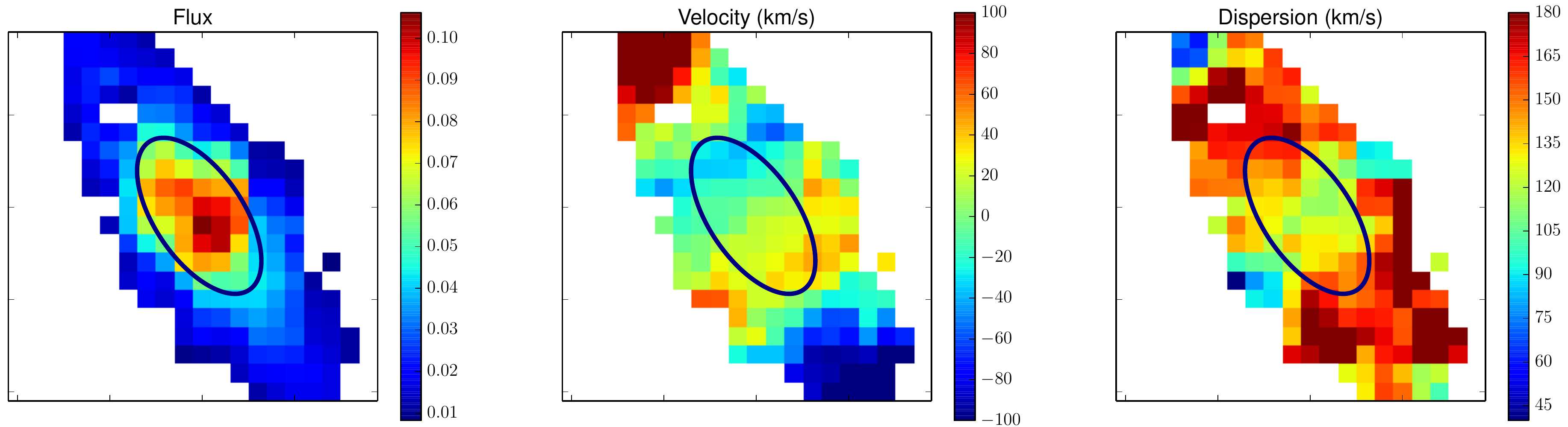}
\caption{This figure shows the derived stellar kinematics for J215634.45-075217.5, with panels as per the top row in Figure \ref{fig:kin_example}. Each panel is 10 $\arcsec$ on a side. This galaxy is a double sigma galaxy, thought to consist of two roughly equal mass counter-rotating disks. The characteristic double-peaked $\sigma$ map (right-hand panel) betrays the true nature of this galaxy.}
\label{fig:kin_2s}
\end{figure}

There is one clear misclassification using the $\lambda_{R}-\epsilon$ diagram which is resolved using the visual classifications. J215634.45-075217.5 is classified as a SR but inhabits a region of the $\lambda_{R}-\epsilon$ not typically populated by true SRs. The kinematics for this galaxy are shown in Figure \ref{fig:kin_2s} and its radial $\lambda_{R}$ profile is shown in black in Figure \ref{fig:radial_lr}. It is visually classified as a FR for a variety of reasons. It is a very oblate system, with an ellipticity of 0.504, the velocity map shows more than one set of alternating regions of approaching and receding velocity and the velocity dispersion map exhibits two peaks on either side of a central depression. These kinematic features are indicative of a so-called double sigma galaxy, initially discovered by \citet{Krajnovic2011} in the ATLAS$^{\rm{3D}}$ sample. They are called double sigma galaxies due to the two characteristics peaks in the velocity dispersion map. These galaxies are thought to consist of two roughly equal-mass counter-rotating disks \citep{Emsellem2011} and are therefore not true SRs. The structure of their stellar kinematic fields, being complex, mimic the low $\lambda_R$ of SRs. For the rest of our analysis we reject this galaxy from our sample, but show its position in plots for completeness.

\subsection{Overall Classifications}

For the remainder of this paper, we will use the $\lambda_{R}-\epsilon$ kinematic classifications for all galaxies. This means that in our sample of 78 galaxies (except J215634.45-075217.5 which we now reject from our sample) we find 13 SRs and 65 FRs. Of these 11 SRs and 62 FRs are cluster members.

\section{Error Analysis}
\label{sec:errors}

As discussed in Section \ref{sec:samidr}, SAMI data cubes are generated from raw, dithered, fibre spectra using a drizzle-like algorithm (Sharp et al. in prep.). This results in data cubes in which the spaxels are not independent: both the data and noise are in a single spaxel are correlated with surrounding spaxels. The variance in a single spaxel is, by definition, correct for that spaxel. However, blindly combining spectra and their corresponding variance spectra neglects the covariance between spaxels and results in an overestimate of the S/N for the combined spectrum. This complicates the error analysis for integrated parameters derived from higher order SAMI data products, such as kinematic maps. In this section we discuss in detail how we deal with this problem, particularly with respect to our calculation of $\lambda_{R}$ and the associated uncertainty.

\subsection{Scaled Kinematic Uncertainties}

We use pPXF \citep{CappellariEmsellem2004} to derive the stellar kinematics for the SAMI Pilot Survey galaxies (see Section \ref{sec:ppxf} and Paper I). This produces spatially resolved maps of flux, V and $\sigma$ along with corresponding uncertainty maps. As discussed above, the error on any individual spaxel in these maps is correct (see Sharp et al. in prep. for more details), but combining these errors directly is invalid since this neglects covariance between spaxels. When calculating uncertainties on integrated properties derived from these maps we must account for covariance between neighbouring spaxels.

To achieve this we wish to scale the uncertainties on flux, V and $\sigma$ by some appropriate factor that accounts for the missing covariance when combining spaxels. To estimate the appropriate scaling factor we use the covariance ``cube'' created by the SAMI cubing code (see Section \ref{sec:samidr}). For each spaxel in each wavelength slice this cube contains a 5 x 5 grid of values showing how correlated the spaxel is with its neighbouring spaxels. The central value in this 5 x 5 grid corresponds to the spaxel itself and therefore always has a value of 1. The sum over this grid shows how correlated the pixel is with all of its neighbours. This total correlation value, when multiplied by the correct variance value, is the simplest way to account for the true uncertainty (variance and covariance) in a single spaxel, when combining that spaxel with all of its neighbours.

\begin{figure*}
\centering
\includegraphics[width=17cm]{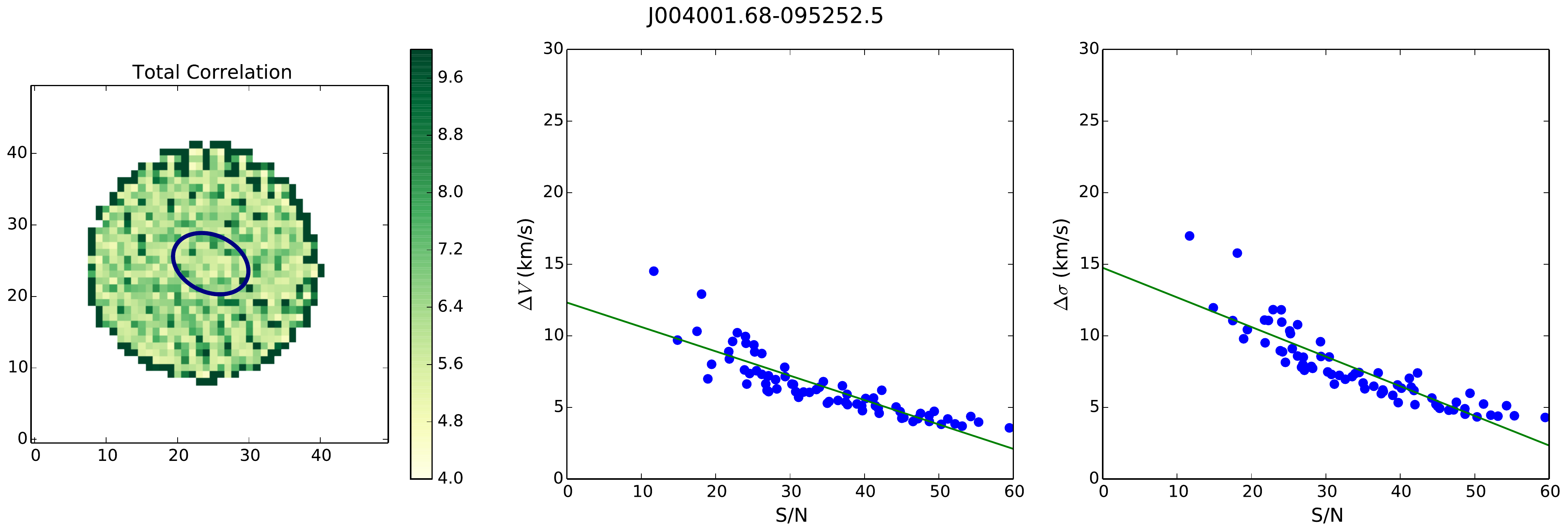}
\caption{The left-hand panel shows the total correlation map for J004001.68-095252.5 averaged across 200 wavelength slices of the total data cube with the blue ellipse representing the effective radius of the galaxy in question. The axes are in pixels with each pixel 0.5$\arcsec$ on a side and so the map is 25$\arcsec$ on a side. The centre and right-hand panel show the trend of the kinematic uncertainties with S/N. The centre panel shows the uncertainties on V and the right-hand panel the uncertainties on $\sigma$.}
\label{fig:cov_all}
\end{figure*}

A map of these values can be generated at each wavelength slice, creating a full cube of total correlation maps. The cube can then be averaged across a wavelength range of interest. The left-hand panel of Figure \ref{fig:cov_all} shows the total correlation map for one of the galaxies in our sample, averaged across 200 wavelength slices in the centre of the data cube. The total correlation factor can vary quite sharply from spaxel to spaxel. This is explained by the way we resample our data. In some cases an output spaxel will fall on the outer edge of the footprint of an input fibre. This spaxel may receive little or no flux from other adjacent fibres and will therefore be strongly correlated with its neighbouring spaxels, leading to a high total correlation factor. In other cases an ouput spaxel may fall in the centre of the footprint of an input fibre. Most of the flux in the input fibre is contributed to that spaxel and so it is only weakly correlated with adjacent spaxels, yielding a low total correlation factor.

To scale the uncertainties on the flux the median total correlation factor, $F_{c}$, is computed within the fiducial radius at which $\lambda_{R}$ is measured. This method overestimates the true value of $F_{c}$ by some negligible amount as it includes correlation between spaxels on the edge of the aperture and nearest neighbours just outside the aperture. This generates a single factor suitable for scaling the variance in the flux at the same wavelength slice(s) at which the correlation factor was measured. We measure the flux over 200 wavelength slices and therefore take the median of 200 wavelength slices to calculate the correlation factor. The square root of $F_{c}$ can be used to scale the uncertainties (sigma) on the flux measurements.

To scale the uncertainties on V and $\sigma$ we first consider the relation between these quantities and the S/N in each spaxel. For reasonably high values of S/N one expects the uncertainties on V and $\sigma$ to scale linearly with S/N. However, for lower S/N this relation breaks down, with a trend towards increasingly large uncertainties.  The centre and right-hand panels in Figure \ref{fig:cov_all} illustrates this. The centre panel shows the uncertainties on V in the spaxels used to calculate $\lambda_{R}$ as a function of S/N in those spaxels. The right-hand panel shows the same plot for the uncertainties on $\sigma$. In each case the green line indicates a linear fit to the points. It is clear that the linear fits do not describe the data perfectly, and in fact a second-order fit may do a better job. However, the linear fit describes the majority of the data points well enough, so we adopt linear scaling between the flux uncertainties and the kinematic uncertainties. The reason for this is to enable us to scale the kinematic uncertainties using the same factor as the flux uncertainties, simplifying our calculation of the uncertainties on $\lambda_{R}$. Thus we scale all three uncertainty maps (flux, V, and $\sigma$) using the square root of $F_{c}$.

\subsection{Uncertainties on $\lambda_R$}
\label{sec:lambda_errors}

We use the scaled uncertainty maps for flux, V and $\sigma$ to calculate the uncertainty in $\lambda_{R}$ for each galaxy in two ways. First, we use the analytic method described by \citet{Houghton2013} in their Appendix A (unlike \citet{Houghton2013} we neglect the covariance between V and $\sigma$, which is very small). The resulting analytic uncertainties on $\lambda_{R}$ are shown in Table \ref{tab:all_gals}, in the column marked $\Delta\lambda_{R}$.

\begin{figure}
\centering
\includegraphics[width=8.5cm]{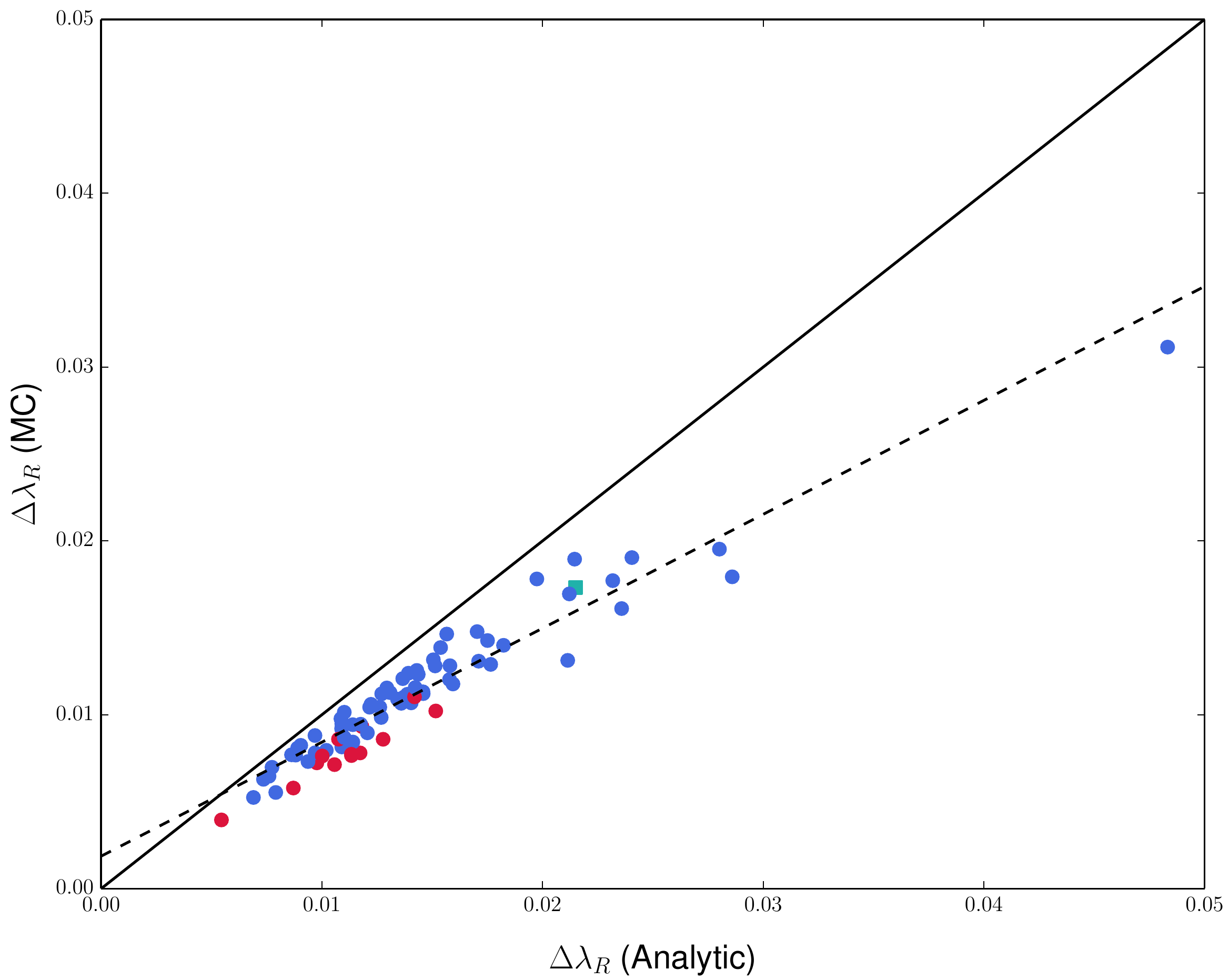}
\caption{This figure compares the uncertainties on $\lambda_{R}$ derived analytically and using a Monte Carlo calculation. The blue points show FRs and the red show SRs. The turquoise square represents the double sigma galaxy discussed in Section \ref{sec:other_class}. The solid black line illustrates a one-to-one correlation and the dashed black line is a fit to the points. The analytic uncertainties are on average 25\% larger than the MC uncertainties and the former values are adopted for the remainder of our analysis.} 
\label{fig:MC}
\end{figure}

In addition to the analytic calculation, we calculate the uncertainty in $\lambda_R$ using a Monte Carlo approach. We use the scaled uncertainty maps for each parameter to generate 1000 instances of the flux, V, and $\sigma$ maps, calculating $\lambda_R$ for each iteration. The spread in the results is a measure of the uncertainty in $\lambda_R$. The resulting uncertainty values for each galaxy are presented in Table \ref{tab:all_gals} in the column marked $\Delta\lambda_{R}$ (MC). Figure \ref{fig:MC} shows the comparison between the analytic and MC uncertainties on $\lambda_{R}$. There is a systematic difference between the two measurements such that the analytic uncertainties are an average of 25\% larger than the MC uncertainties, though the discrepancy is smaller for smaller uncertainty values. For the analysis presented here we adopt the larger of the two - the analytic uncertainties.

\subsection{Uncertainties on $f_{SR}$}

Uncertainties on $f_{SR}$ are derived using simple binomial error calculation. The uncertainties are reported in Table \ref{tab:cluster_members}.

\section{Results and Discussion}
\label{sec:results}

\subsection{Slow Rotator Fraction}

\begin{figure*}
\centering
\includegraphics[width=12cm]{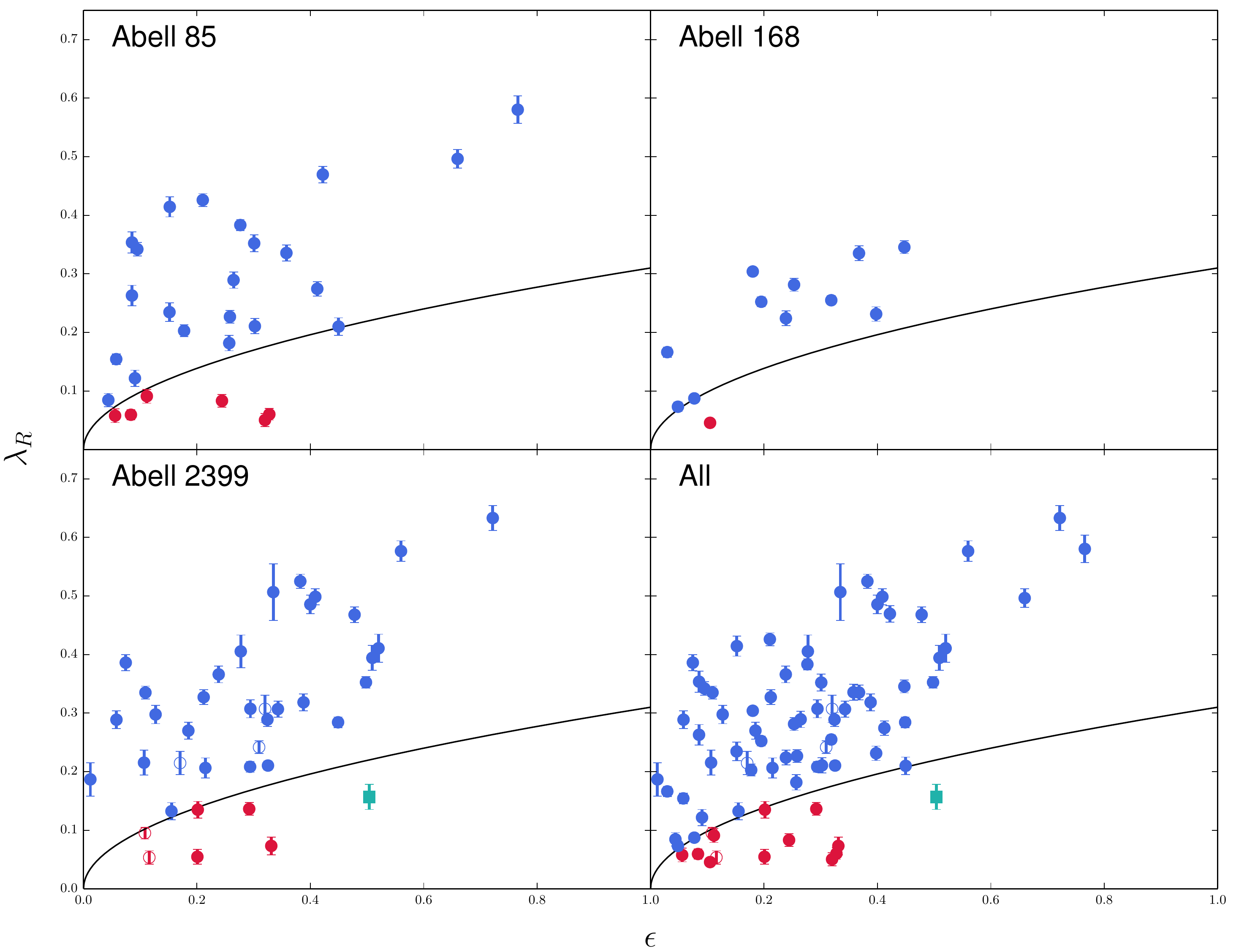}
\caption{The $\lambda_{R}-\epsilon$ diagram for the 79 ETGs in the SAMI Pilot survey. The top-left panel shows galaxies in Abell 85, top-right Abell 168 and bottom-left Abell 2399. The bottom-right panel shows the entire sample. In all panels the black line shows the division between FRs and SRs defined at R$_e$. The red points are SRs and blue points are FRs. Filled symbols represent cluster members, as discussed in Section \ref{sec:clustermembership}, whereas open symbols represent foreground or background galaxies. The turquoise square point is a double sigma galaxy and is discussed in Section \ref{sec:other_class}.}
\label{fig:lambda_clusters}
\end{figure*}

Using the quantitative FR/SR classifications we plot the $\lambda_{R}-\epsilon$ diagram for all 79 ETGs in the three clusters in Figure \ref{fig:lambda_clusters}. For compactness we plot $\lambda_{R}$ and $\epsilon$ as measured for each galaxy on the same plot, even though these quantities have been measured at different fiducial radii for different objects. The colour-coding is according to the $\lambda_{R}-\epsilon$ classifications using three different dividing lines (dependent on the choice of fiducial radius, see Figure \ref{fig:lambda_multiline}), but only the line defined at R$_e$ is shown on the plots to guide the eye. Closed symbols represent cluster members, as discussed in Section \ref{sec:clustermembership}, and open symbols represent foreground or background galaxies (non-members).   

From here on this paper will focus on cluster member ETGs only, and observed non-members are shown on plots only for completeness. We find slow rotators in each of the three cluster samples examined. The overall number of SRs found in each cluster is 6, 1 and 4 for Abell 85, Abell 168 and Abell 2399 respectively. We calculate the SR fraction, $f_{SR}$, for each of our clusters and for the entire sample of cluster member ETGs. $f_{SR}$, is defined as:

\begin{equation}
f_{SR}=\rm{\frac{N(SR)}{N(FR)+N(SR)}}
\end{equation}

\noindent where N(SR) is the number of SRs observed and N(FR) the number of FRs observed in the sample of ETG cluster members. The SR fractions for our sample, both total and split by cluster, are presented in Table \ref{tab:cluster_members}. They are $0.21\pm0.08$, $0.08\pm0.08$ and $0.12\pm0.06$ for Abell 85, Abell 168 and Abell 2399 respectively, with an overall fraction of $0.15\pm0.04$ for the entire sample. These values are consistent with each other, given the low numbers of galaxies in the individual clusters. They are also consistent with previous studies, which have shown a remarkably steady SR fraction, about 0.15, across many different GHEs, from the field to very massive clusters \citep{Houghton2013}. 

\begin{table}
\centering
\begin{tabular}{|c|c|c|c|}
\hline
Cluster & Member ETGs & SRs & $f_{SR}$\\
\hline
Abell 85 & 28 & 6 & $0.21\pm0.08$ \\
Abell 168 & 12 & 1 & $0.08\pm0.08$ \\
Abell 2399 & 33$^*$ & 4 & $0.12\pm0.06$ \\
\hline
Total & 73$^*$ & 11 & $0.15\pm0.04$ \\ 
\hline
\end{tabular}
\caption{Slow rotator fractions in the three clusters observed in the SAMI Pilot Survey. The absolute numbers of cluster member ETGs (column 2) and of SRs (column 3) are given along with the fraction, $f_{SR}$. The uncertainties on $f_{SR}$ are calculated using a simple binomial framework. $^*$ denotes that this number has been calculated excluding the double sigma galaxy J215634.45-075217.5.}
\label{tab:cluster_members}
\end{table}

\subsection{The Kinematic Morphology-Density Relation}
\label{sec:kinmorphdens}

\begin{figure*}
\centering
\includegraphics[width=11cm]{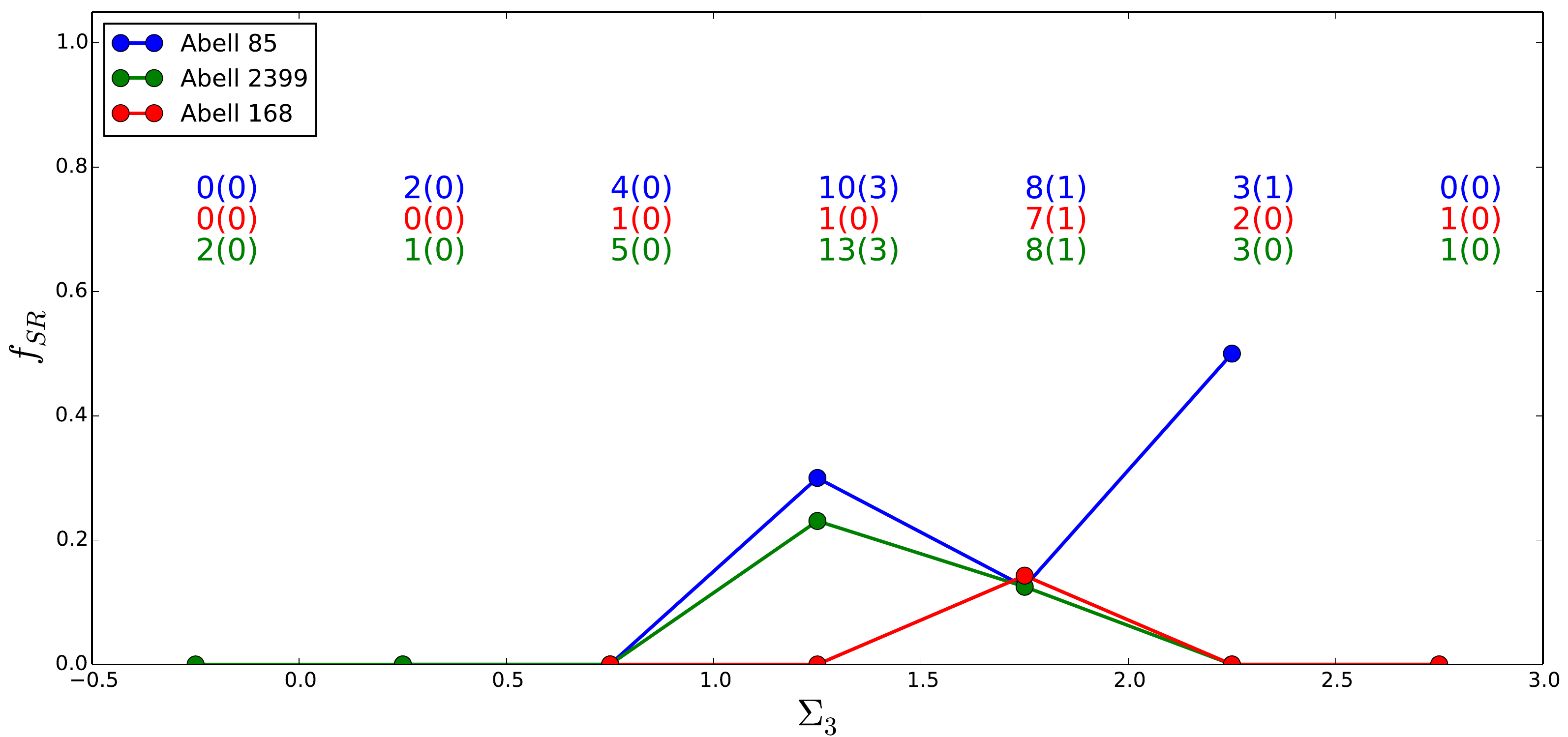}
\caption{The kinematic morphology-density relation for all three clusters. The galaxy number density, $\Sigma_{3}$ as described in Section \ref{sec:lpe}, is on the x-axis with $f_{SR}$ shown on the y-axis. The blue points show the relation for Abell 85, the red for Abell 168 and the green for Abell 2399. The coloured numbers correspond to the total number of ETGs in each bin for the corresponding galaxy cluster. Abell 85 shows an increase $f_{SR}$ towards the densest environments in the cluster. However this trend is not seen in Abell 168 or Abell 2399.
\label{fig:morph_dens}}
\end{figure*}

To investigate the kinematic morphology-density relation we calculate $f_{SR}$ as a function of LPE. Here we use the 3rd nearest neighbour measurements ($\Sigma_3$) as discussed in Section \ref{sec:lpe}. In Figure \ref{fig:morph_dens} we show the resulting relations for all three of our clusters, with Abell 85 shown in blue, Abell 168 in red and Abell 2399 in green. The number of cluster member ETGs in each bin is shown at the top of the plot in the appropriate colour.

In Abell 85 we see that $f_{SR}$ increases at high densities. However, we also note that there are high values of $f_{SR}$ in lower/intermediate density bins. This has not been seen before and implies that there are SRs distributed across Abell 85, not just at the centre of the cluster. Abell 85 is discussed in detail in Section \ref{sec:a85}. 

In Abell 168 we see that $f_{SR}$ peaks in an intermediate density region of the cluster but with a decrease towards the densest region. However as there is only one SR in this cluster it is not possible to make firm conclusions for this cluster. We discuss what we see in this cluster in more deatil in \ref{sec:a168}.

In Abell 2399 we also see a decrease in $f_{SR}$ towards the densest region of the cluster, with higher values of $f_{SR}$ in lower/intermediate density bins. This suggests that the SRs in Abell 2399 do not preferentially lie in the densest part of the cluster but instead at lower/intermediate densities. In fact, both central galaxies in Abell 2399 (the BCG and a bright companion) are classified as FRs. This is discussed more in Section \ref{sec:a2399}.

To fully understand the trends we see in the kinematic morphology-density relation it is necessary to examine the on-sky distribution of the FR and SR populations within the three clusters. The left-hand panels of Figures \ref{fig:a85_all}, \ref{fig:a168_all} and \ref{fig:a2399_all} show the projected on-sky positions of the observed sample of ETGs for each of the three clusters. In each of these Figures the FRs are plotted in blue and the SRs in red, with closed symbols representing cluster members and open symbols representing non-members. The contours show galaxy density as calculated using the SAMI Cluster Redshift Data described in Section \ref{sec:clustermembership} (Owers et al. in prep.). The dashed red circles show R$_{200}$ for each cluster.

The left-hand panel of Figure \ref{fig:a168_all} shows the on-sky distribution of ETGs in Abell 168 and we see that the only slow rotator in the cluster lies close to but not at the centre of the cluster (it is in fact the BCG). The densest region in this cluster is offset both from the BCG and from the centre position of the cluster. This clearly reflects the trend seen in the kinematic morphology-density relation in Figure \ref{fig:morph_dens}. 

For both Abell 85 and Abell 2399 (the left-hand panels of Figures \ref{fig:a85_all} and \ref{fig:a2399_all} respectively) the distribution of SRs is quite spread out. We see SRs on the outskirts of both of these clusters and we hypothesize that these SRs could be associated with in-falling groups and are joining the cluster for the first time. We discuss each cluster in more detail, highlighting some interesting galaxies, in the following Sections.

\subsection{Abell 85}
\label{sec:a85}

\begin{figure*}
\centering
\includegraphics[width=17cm]{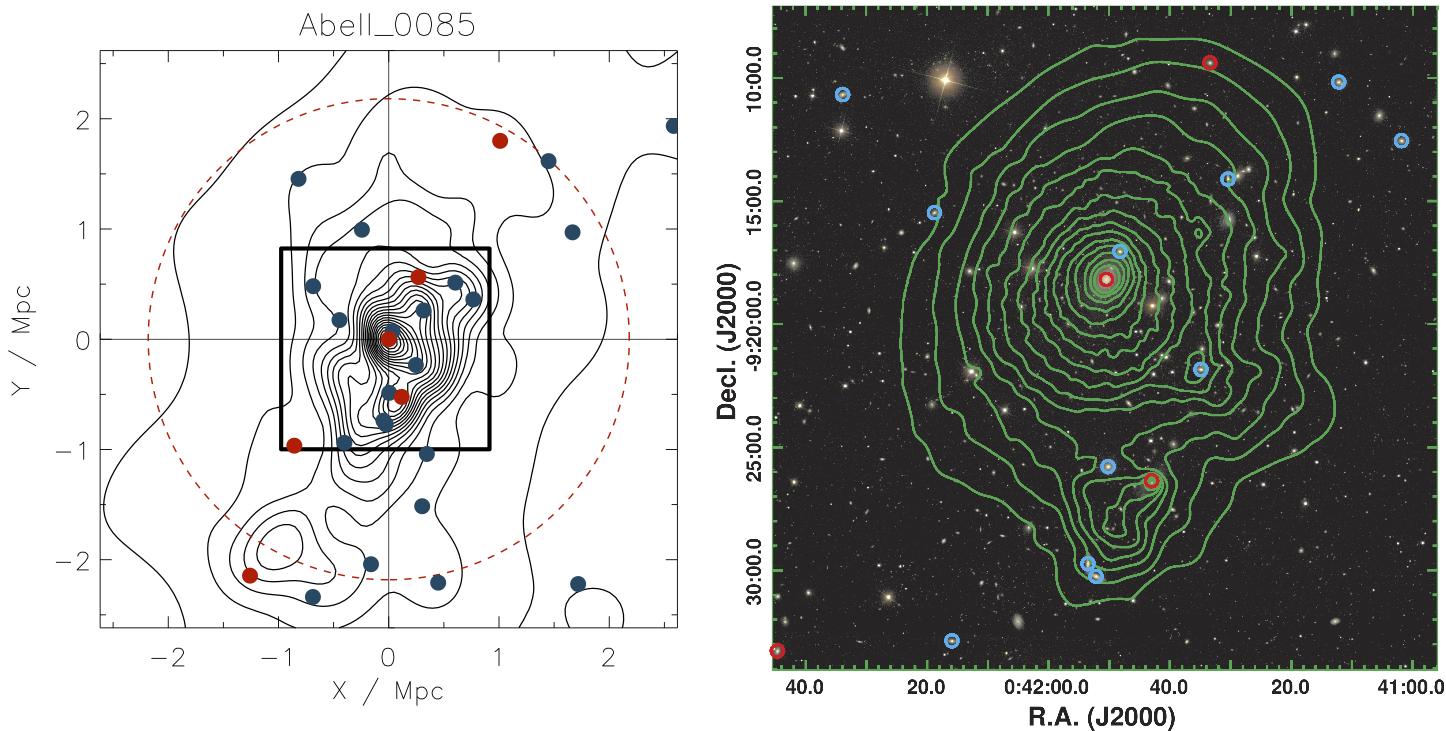}
\caption{The left-hand panel shows the on-sky distribution of ETGs observed with SAMI in Abell 85. The red symbols show SRs and the blue FRs. The red dashed circle indicates R$_{200}$ and the black dotted circle has a one-degree in radius, matching the initial SAMI Pilot sample selection. The black contours show galaxy density. The right-hand panel shows the central part of Abell 85, as indicated by the black box in the left-hand panel. The image is a $gri$-colour composite from SDSS, with {\it Chandra} X-ray contours overlaid in green. The contours are logarithmically spaced with the faintest indicating 5\,x\,$10^9$ photons\,cm$^{-2}$\,s$^{-1}$\,arcsec$^{-2}$ and the brightest 650\,x\,$10^9$ photons\,cm$^{-2}$\,s$^{-1}$\,arcsec$^{-2}$. The red circles indicate SRs and the blue show FRs, according to the SAMI quantitative classifications. Only cluster members are indicated in this figure. The large SR in the centre of the main concentration of X-ray emission is the BCG for this cluster, J004150.46-091811.2. The SR to the south of the main concentration of X-ray emission appears to be in-falling and is perhaps the central galaxy in an in-falling group.}
\label{fig:a85_all}
\end{figure*}

Abell~85 is the most massive of our three clusters, with a mass of $11.9\pm1.4$ x $10^{14}\rm{M}_{\odot}$. Previous analysis of ROSAT PSPC and {\it XMM-Newton} observations of this cluster have found that it has undergone merging events in the distant past and is still potentially undergoing merging with various groups today \citep{Durret1998, Durret2005}. Thus this cluster is not fully relaxed. Furthermore, \citet{Kempner2002} used the {\it Chandra} observations presented here to confirm that the so-called ``southern blob'', clearly seen to the south of the cluster centre in the right-hand panel of Figure \ref{fig:a85_all}, is merging with the main cluster.

Abell 85 harbours six SRs. A $gri$-colour composite image of the central 30$\arcmin$ of the cluster is shown in the right-hand panel of Figure \ref{fig:a85_all} with {\it Chandra} contours overlaid. The left-hand panel shows the on-sky distribution of galaxies. It is clear that the SRs in Abell~85 are found across the cluster and are not necessarily concentrated in the central parts. This supports the hypothesis put forward in \citet{Houghton2013} that SRs are not preferentially formed at the centres of clusters, but are formed with a constant ratio to FRs across a large range of GHEs. This is inferred from the fact that $f_{SR}$, the fraction of SRs in the ETG population, remains constant across those GHEs already investigated, including the clusters in this paper.

If  SRs are indeed formed with equal efficiency across a range of GHEs then we should see some evidence for SR formation in groups. This study did not cover any isolated groups. However, the X-ray emission in Abell~85 strongly suggests that we do see an in-falling group, the southern blob, close to the BCG. The main concentration of X-ray emission in Abell~85 is centred on J004150.46-091811.2, the BCG, classified by SAMI as a SR (see the right hand panel of Figure \ref{fig:a85_all}). Unsurprisingly, this galaxy lives in the densest part of Abell~85, the rough centre of the cluster. The SAMI stellar kinematics for the BCG are shown in the top row of Figure \ref{fig:all_centrals}. To the south-west of the BCG there lies another, less extended, concentration of X-ray emission, the southern blob. This blob is coincident with another SR, J004143.00-092621.9 and with a local galaxy over-density (see the left-hand panel of Figure \ref{fig:a85_all}), suggesting that the X-ray emission is from a galaxy group falling into Abell~85. This is supported by the morphology of the X-ray emission, which shows a definite elongation, reminiscent of a tail. In addition, this galaxy and the neighbouring FR, J004150.17-092547.4 to the east both exhibit a head-tail structure in 90cm radio emission (see \citet{Kempner2002} and references therein), a feature very suggestive of infall. 

The SR (J004143.00-092621.9) is therefore likely to be the central galaxy of the in-falling group. It most likely formed outside the cluster and is now being accreted along with the rest of the group. The kinematics for J004143.00-092621.9 are shown in the second row of Figure \ref{fig:all_centrals}. This galaxy is a very clear SR with agreement between the quantitative and qualitative classifications. This is evidence for SR formation within a group which is subsequently ingested into a large cluster.

\subsection{Abell 168}
\label{sec:a168}

\begin{figure*}
\centering
\includegraphics[width=17cm]{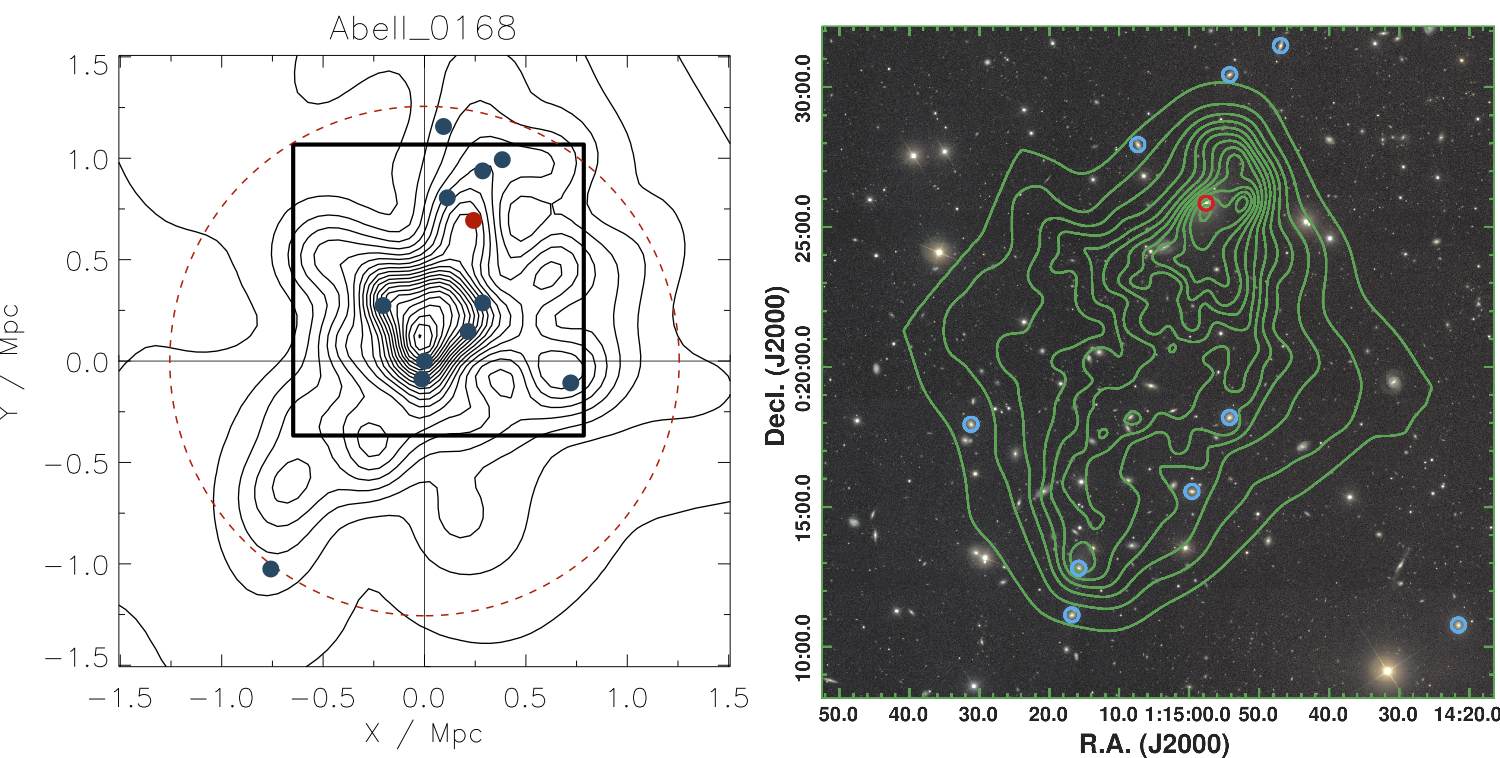}
\caption{Abell 168. The panels and symbols are as for Figure \ref{fig:a85_all} except that in this case the contours are linearly spaced with the faintest showing 2\,x\,$10^9$ photons\,cm$^{-2}$\,s$^{-1}$\,arcsec$^{-2}$ and the brightest showing 13\,x\,$10^9$ photons\,cm$^{-2}$\,s$^{-1}$\,arcsec$^{-2}$. This cluster is a well-known cluster-cluster merger and as well as exhibiting two clear peaks in galaxy density (left-hand panel) with two X-ray peaks visible in the {\it Chandra} contours.}
\label{fig:a168_all}
\end{figure*}

Abell 168 is a well-known cluster-cluster merger \citep{Ulmer1992, Yang2004}. \citet{Hallman2004} used the {\it Chandra} observations presented here to determine that the merger is in a late stage, with the two sub-clusters having already passed one another and now on trajectory for a second pass. \citet{HwangLee2009} find an increased fraction of active galaxies in the region between the two merging subclusters suggesting that the first pass in the merging process triggered some activity.

The right-hand panel of Figure \ref{fig:a168_all} shows a SDSS $gri$-colour composite image of the central 20' of Abell 168, with overlaid {\it Chandra} contours in green. There are two main peaks of X-ray emission in the central pars of the cluster corresponding to the two progenitor clusters which are now merging. The strongest of these emission peaks is the northern one and it is coincident with the BCG, J011457.59+002550.8, which is classified as a SR using the SAMI kinematics (this is the only SR found in this cluster and its kinematics are shown in the third row of Figure \ref{fig:all_centrals}). We cannot say whether this galaxy formed in the centre of the progenitor cluster it occupies or whether it migrated there through some process such as dynamical friction.

To the south-east of the BCG the second X-ray emission peak is coincident with J011515.78+001248.4. This galaxy is classified as a FR and its kinematics are shown in the fourth row of Figure \ref{fig:all_centrals}. It is likely the central galaxy of the second progenitor cluster now merging. Although this galaxy is classified as a FR it is clearly very close to the dividing line in $\lambda_{R}-\epsilon$ space and the visual classification of this galaxy nominates it as a SR candidate - in disagreement with the quantitative classification. This galaxy is round and has high mass, low rotation and high velocity dispersion. There is a chance it could in fact be a SR. Either way, this object dominates the lower of the two X-ray emission peaks.

In this cluster the density peak is not coincident with either of the X-ray peaks corresponding to the two progenitor clusters, and in fact the densest region in the cluster lies between the two X-ray peaks. This explains why the kinematic morphology density relation in Abell 168 does not peak in the densest region of the cluster. This does not contradict the prevailing picture of SR formation occuring in high density regions, since the SR in this cluster could easily have formed before the merging event, in what was likely the densest part of the progenitor cluster.

\subsection{Abell 2399}
\label{sec:a2399}

\begin{figure*}
\centering
\includegraphics[width=17cm]{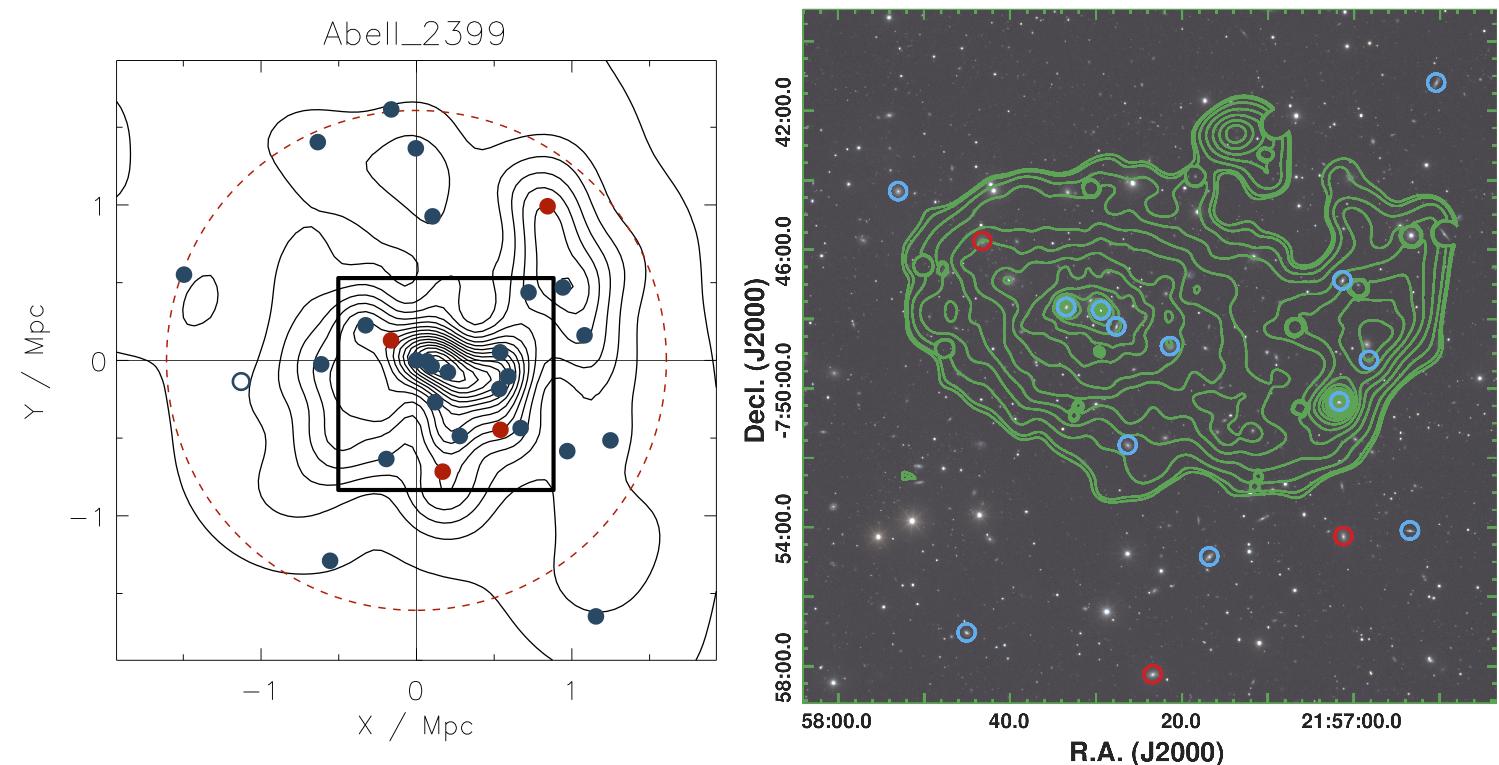}
\caption{Abell 2399. The symbols are as for Figure \ref{fig:a85_all}, except that in this case the X-ray contours in the right-hand panel show {\it XMM-Newton} data and are in counts. The contours are square-root spaced with the faintest showing 6.4\,x\,$10^9$ counts\,cm$^{-2}$\,s$^{-1}$\,arcsec$^{-2}$ and the brightest showing 48\,x\,$10^9$ counts\,cm$^{-2}$\,s$^{-1}$\,arcsec$^{-2}$. The two FRs coincident with the central X-ray emission concentration are the brightest galaxies in the cluster, including the BCG. In the left-hand panel open symbols indicate cluster non-members.}
\label{fig:a2399_all}
\end{figure*}

Abell 2399 has previously been observed with {\it XMM-Newton} as part of the REXCESS Survey \citep{Bohringer2007}. \citet{Bohringer2010} found that Abell 2399 shows a bimodal structure in X-rays. Abell 2399 was also studied in the optical as part of the WINGS survey. However, they did not detect substructure in the cluster from galaxy clustering only \citep{Ramella2007}. Nonetheless, Abell 2399 is likely to be a cluster-cluster merger and is the most complicated of the clusters we observed. Unlike Abell 168 the density peak in this cluster is coincident with both the peak in X-ray emission and the location of the BCG.

\begin{figure*}
\centering
\includegraphics[width=17cm]{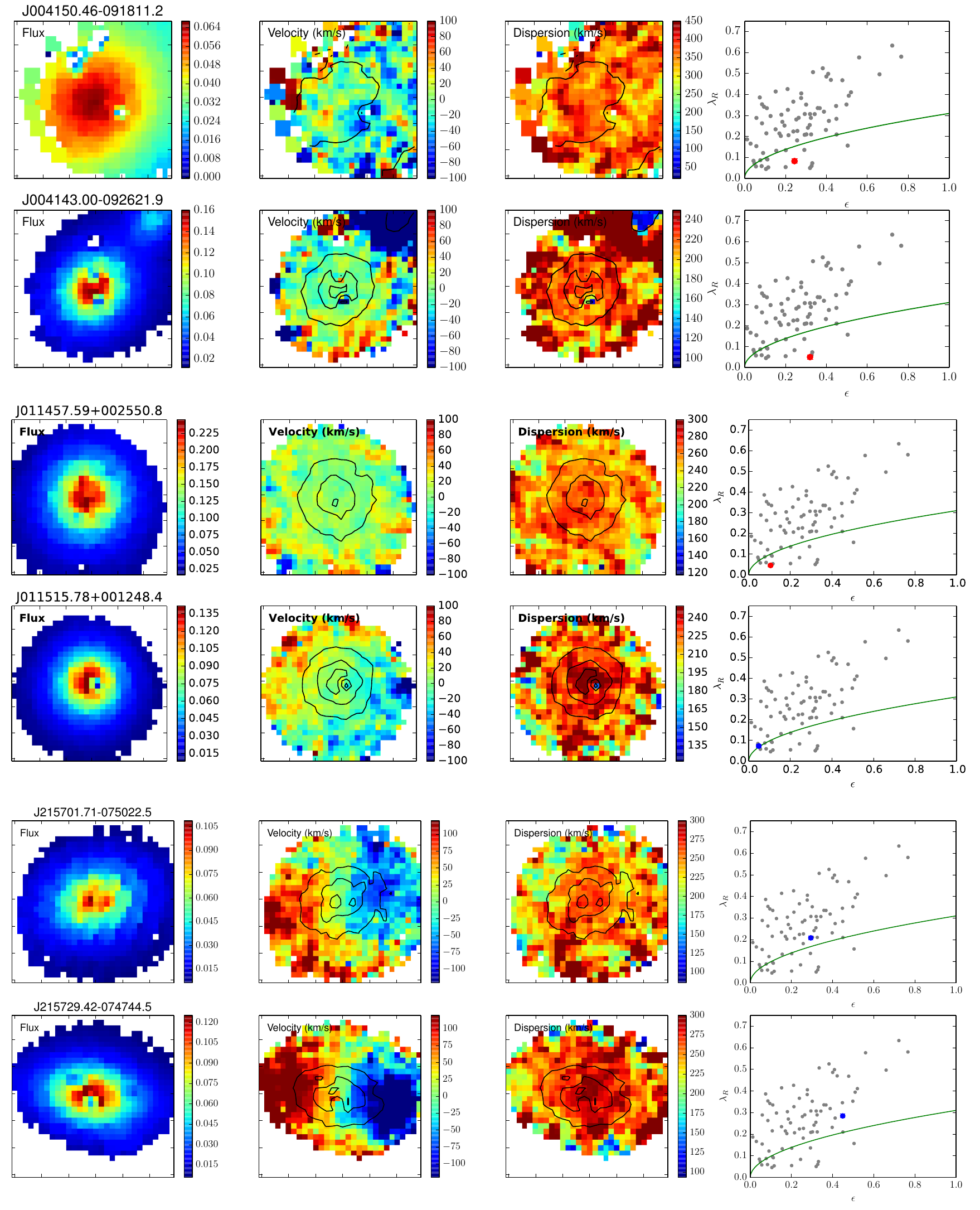}
\caption{Selected interesting galaxies from each of the three clusters. Each row displays information for a single galaxy, wherein the first three panels show the SAMI kinematic data (flux, V and $\sigma$) with the SAMI flux contours overlaid in black on the latter two maps. The fourth panel in each row shows the $\lambda_{R}-\epsilon$ diagram for the entire sample with the galaxy of interest highlighted. The first two rows show the J005150.46-091811.2 the BCG in Abell 85 and J004143.00-092621.9 the central SR in a group falling into Abell 85 (see text for details). The second two rows show J011457.59+002550.8, the BCG of Abell 168 and J011515.78+001248.4 another galaxy of interest in Abell 168. The last two rows show J215701.71-075022.5 the BCG of Abell 2399 and a clear FR and J215729.42-074744.5, a close companion of the BCG, also a clear FR.}
\label{fig:all_centrals}
\end{figure*}

We observe four SRs in Abell 2399 but they are not found close to the centre of the cluster. The right-hand panel of Figure \ref{fig:a2399_all} shows a SDSS $gri$-colour composite image of the central 20' of Abell 2399, overlaid with X-ray contours from {\it XMM-Newton}. The central galaxies in this cluster are all FRs. Most notably, the BCG (J215701.71-075022.5) is a clear FR. This galaxy and its near neighbour (J215729.42-074744.5, also a FR) occupy the X-ray centre of the cluster. The SAMI kinematics of the BCG and its neighbour are shown in the fifth and sixth rows Figure \ref{fig:all_centrals}. 

Since the favoured formation mechanisms for SRs involve multiple minor merging events \citep{Khochfar2011, Naab2013} it is not unreasonable to expect that BCGs, galaxies that are likely to undergo multiple merger events in their histories by virtue of their position at the centre of a potential well, should preferentially be SRs. Previous studies by \citet{Brough2011} and \citet{Jimmy2013} have shown that 70\% of BCGs are indeed SRs. Here we have studied three clusters and found one fast rotating BCG in Abell 2399. Our sample is small but this fraction is consistent with the work of \citet{Brough2011} and \citet{Jimmy2013}. The full SAMI Galaxy Survey will observe eight separate clusters and will provide the perfect sample to investigate this further.

\subsection{Discussion}

We find six, one and four SRs in Abell 85, Abell 168 and Abell 2399. In these clusters we studied  28, 12 and 33 member ETGs, and we therefore measure a roughly constant value of $f_{SR}$ within each cluster. The values measured are consistent with the literature value of 0.15, implying that the mechanism which creates SRs is equally efficient across the many GHEs studied to date \citep{Houghton2013}. However, within a single GHE, the distribution of galaxies with LPE - i.e. the kinematic morphology-density relation - generally shows a trend such that SRs are found at the densest LPE \citep{Cappellari2011, DEugenio2013, Houghton2013}. This could be evidence that the kinematic morphology-density relation is a result of the fact that SRs tend to be high mass systems, which are funnelled to the centres of clusters through dynamical friction. However, Scott et al. (2014) show, using mass-matched samples of FRs and SRs, that the morphology-density relation in Virgo is not caused by dynamical friction alone. In the work presented in this paper we see an increase in $f_{SR}$ in the centres of two out of three of our clusters. However, we also see clear evidence for SRs in low/intermediate density regions in the clusters (i.e. on the cluster outskirts) and can clearly associate one of these with an in-falling group.

Semi-analytic models by \citet{Khochfar2011} find that SRs have more major mergers in their formation histories than FRs, but that they also accrete significantly more mass from multiple dry minor merger events. Cosmological zoom simulations presented by \citet{Naab2013} show a suite of formation scenarios for SRs, with large round dispersion-dominated systems forming by multiple dry minor mergers alone. These SR formation scenarios seem most likely to occur when a central galaxy in a halo accretes many smaller satellites on random trajectories, making central galaxies are more likely to become SRs than non-centrals and partly explaining the observed kinematic morphology-density relation. However, not every halo in the Universe is dominated by a single SR, and we do see SRs on the outskirts of large galaxy clusters, for example in Abell 85 and Abell 2399. There are two other factors to consider. A central galaxy must have access to enough satellites to accrete a large fraction of mass through multiple dry minor mergers, destroying the angular momentum of the progenitor. Equally, the central galaxy must remain central long enough to accrete those satellites. This suggests that there will be a low halo mass cut-off below which it will be difficult for a central galaxy to accrete enough mass by minor mergers to become a SR. 

This can explain the observed kinematic morphology-density relation as well as the presence of SRs on the outskirts of galaxy clusters. If SRs form as the central galaxies of high mass halos (e.g. a high-mass group halo) and then fall into a cluster the high mass group halo is likely to migrate to the centre of the cluster through dynamical friction. As Scott et al. (2014) point out, however, a FR of similar stellar mass is likely to be associated with a lower mass group halo since FRs grow more of their mass through in situ star formation as opposed to minor mergers with multiple satellites. Therefore the FR will not make its way to the centre of the cluster as efficiently as the SR. Thus a combination of SRs tending to form central to their halos and the effects of dynamical friction on infalling group halos explains the fact that we have to date preferentially found SRs at cluster centres.

In the future we plan to extend this study to include more galaxies in these three clusters and to study five more clusters. The cluster sample forms 30\% of the SAMI Galaxy Survey sample and will be selected from the SAMI Cluster Redshift Survey (Owers et al. in prep.). The remaining 70\% of galaxies in the SAMI Galaxy Survey sample are selected from the field and group sample of the Galaxy and Mass Assembly (GAMA; \citet{Driver2011}) survey (for details on the field and group sample selection for the SAMI Galaxy Survey see Bryant et al. in prep.). This sample is ideal to explore the formation of ETGs in groups. We will use the SAMI Galaxy Survey to study both FRs and SRs in clusters, groups, and the field and determine their relationship with their environment and how they evolve. A crucial aspect of this study will be a robust comparison with custom simulations.

\section{Conclusions}
\label{sec:conclusions}

We have examined a sample of 79 ETGs in three cluster regions, 74 of which were cluster members. We classify these objects as fast or slow rotators on the basis of their angular momentum, measured from spatially resolved stellar kinematics. We use $\lambda_R$, a proxy for stellar angular momentum \citep{Emsellem2007}. 

We present three main results in this paper: (i) we find the fraction of SRs, $f_{SR}$ in the ETG population to be $0.21\pm0.08$ (6 SRs in a sample of 28), $0.08\pm0.08$ (1 SR in a sample of 12) and $0.12\pm0.06$ (4 SRs in a sample of 33) for Abell 85, Abell 168 and Abell 2399 respectively and a total fraction of $0.15\pm0.04$ (11 SRs in a fraction of 73) for the entire sample. This is in broad agreement with \citet{Houghton2013} who find that $f_{SR}$ is remarkably constant at $\sim 0.15$ across many GHEs, from the ATLAS$^{\rm{3D}}$ field/group sample to the massive, dense cluster Abell 1689. 

(ii) we present $f_{SR}$ for each of the three galaxy clusters in our sample as a function of galaxy density (local point environment, LPE, see Section \ref{sec:lpe}). This is the kinematic morphology-density relation. In Abell 85 we find that $f_{SR}$ increases towards the densest LPE. However, for Abell 168 and Abell 2399 we do not see such a trend. In the case of Abell 168 this is explained by the nature of the cluster - a merger in which the BCG (the only confirmed SR) is not coincident with the density peak of the cluster. For Abell 2399 the picture is more complicated and indeed we find that the BCG of Abell 2399 is a FR.

(iii) we see SRs at low and intermediate LPE within two of the clusters in our sample, Abell 85 and Abell 2399. These SRs could be central objects associated with in-falling groups, or they could have formed in situ in the cluster. We favour the former explanation and posit that SRs form preferentially as central galaxies of high-mass halos (for example high-mass groups). The high-mass halos are then accreted by galaxy clusters and funneled to the cluster centre through dynamical friction. This process favours SRs over FRs of similar stellar mass as they will, in general, be associated with higher mass halos.

\section*{ACKNOWLEDGEMENTS} 

The SAMI Galaxy Survey is based on observation made at the Anglo-Australian Telescope. The Sydney-AAO Multi-object Integral field spectrograph (SAMI) was developed jointly by the University of Sydney and the Australian Astronomical Observatory. The SAMI input catalogue is based on data taken from the Sloan Digital Sky Survey, the GAMA Survey and the VST ATLAS Survey. The SAMI Galaxy Survey is funded by the Australian Research Council Centre of Excellence for All-sky Astrophysics (CAASTRO), through project number CE110001020, and other participating institutions. The SAMI Galaxy Survey website is http://sami-survey.org/ 

MSO acknowledges the funding support from the Australian Research Council through a Super Science Fellowship (ARC FS110200023)

SMC acknowledges the support of an ARC future fellowship (FT100100457).

This work was supported by the Astrophysics at Oxford grants (ST/H002456/1 and ST/K00106X/1) as well as visitors grant (ST/H504862/1) from the UK Science and Technology Facilities Council. RLD acknowledges travel and computer grants from Christ Church, Oxford. RLD is also grateful for support from the Australian Astronomical Observatory Distinguished Visitors programme, the ARC Centre of Excellence for All Sky Astrophysics, and the University of Sydney during a sabbatical visit.

JTA acknowledges the award of an ARC Super Science Fellowship (FS110200013).

ISK is the recipient of a John Stocker Postdoctoral Fellowship from the Science and Industry Endowment Fund (Australia).

LC acknowledges support under the Australian Research CouncilÕs Discovery Projects funding scheme 
(project number 130100664).

This research made use of Montage, funded by the National Aeronautics and Space Administration's Earth Science Technology Office, Computational Technologies Project, under Cooperative Agreement Number NCC5-626 between NASA and the California Institute of Technology. The code is maintained by the NASA/IPAC Infrared Science Archive.

This research made use of Astropy, a community-developed core Python package for Astronomy (Astropy Collaboration, 2013)

\bibliography{SAMI_StellarKin}
\bibliographystyle{mn2e}

\appendix

\section[]{SAMI Pilot Survey Galaxy Properties.}

\begin{table*}
\centering
\begin{tabular}{|c|c|c|c|c|c|c|c|c|c|}
\hline
Galaxy Name & Cluster & Membership & Ellipticity & Re & $\lambda_R$ & $\Delta\lambda_R$ & $\Delta\lambda_{R}$ & Kinematic & Fiducial \\
& & & & ($''$ Major Axis) & & & (MC) & Class & Radius\\
\hline
J003906.77-084758.3 & ABELL0085 & 1 & 0.09 & 2.54 & 0.263 & 0.018 & 0.014 & FR & Re\\
J004001.68-095252.5 & ABELL0085 & 1 & 0.28 & 2.76 & 0.383 & 0.01 & 0.009 & FR & Re\\
J004004.88-090302.6 & ABELL0085 & 1 & 0.36 & 4.24 & 0.336 & 0.014 & 0.011 & FR & Re\\
J004018.68-085257.1 & ABELL0085 & 1 & 0.45 & 3.81 & 0.21 & 0.015 & 0.013 & FR & Re\\
J004046.47-085005.0 & ABELL0085 & 1 & 0.06 & 3.97 & 0.058 & 0.012 & 0.008 & SR & Re\\
J004101.87-091233.1 & ABELL0085 & 1 & 0.15 & 3.46 & 0.235 & 0.016 & 0.013 & FR & Re\\
J004112.21-091010.2 & ABELL0085 & 1 & 0.06 & 2.24 & 0.154 & 0.009 & 0.008 & FR & Re\\
J004122.06-095240.8 & ABELL0085 & 1 & 0.41 & 8.87 & 0.274 & 0.012 & 0.01 & FR & Re/2\\
J004128.56-093426.7 & ABELL0085 & 1 & 0.42 & 4.42 & 0.47 & 0.014 & 0.011 & FR & Re\\
J004130.42-091406.7 & ABELL0085 & 1 & 0.3 & 2.03 & 0.211 & 0.013 & 0.012 & FR & Re\\
J004131.25-094151.0 & ABELL0085 & 1 & 0.09 & 4.26 & 0.354 & 0.018 & 0.014 & FR & Re\\
J004133.41-090923.4 & ABELL0085 & 1 & 0.11 & 2.38 & 0.091 & 0.012 & 0.009 & SR & Re\\
J004134.89-092150.5 & ABELL0085 & 1 & 0.26 & 2.29 & 0.182 & 0.013 & 0.011 & FR & Re\\
J004143.00-092621.9 & ABELL0085 & 1 & 0.32 & 8.12 & 0.05 & 0.011 & 0.008 & SR & Re/2\\
J004148.22-091703.1 & ABELL0085 & 1 & 0.26 & 2.59 & 0.227 & 0.011 & 0.01 & FR & Re\\
J004150.17-092547.4 & ABELL0085 & 1 & 0.18 & 4.66 & 0.203 & 0.01 & 0.008 & FR & Re\\
J004150.46-091811.2 & ABELL0085 & 1 & 0.24 & 16.34 & 0.083 & 0.011 & 0.007 & SR & Re/2\\
J004152.16-093014.8 & ABELL0085 & 1 & 0.21 & 4.67 & 0.426 & 0.011 & 0.008 & FR & Re\\
J004153.50-092943.9 & ABELL0085 & 1 & 0.66 & 7.74 & 0.496 & 0.016 & 0.012 & FR & Re/2\\
J004200.64-095004.0 & ABELL0085 & 1 & 0.77 & 7.0 & 0.58 & 0.024 & 0.016 & FR & Re\\
J004205.86-090240.7 & ABELL0085 & 1 & 0.09 & 3.02 & 0.122 & 0.014 & 0.011 & FR & Re\\
J004215.91-093252.0 & ABELL0085 & 1 & 0.3 & 2.62 & 0.352 & 0.015 & 0.011 & FR & Re\\
J004218.75-091528.4 & ABELL0085 & 1 & 0.26 & 3.03 & 0.289 & 0.014 & 0.012 & FR & Re\\
J004233.86-091040.5 & ABELL0085 & 1 & 0.09 & 3.37 & 0.342 & 0.011 & 0.009 & FR & Re\\
J004233.99-095442.2 & ABELL0085 & 1 & 0.04 & 3.25 & 0.084 & 0.011 & 0.008 & FR & Re\\
J004242.26-085528.1 & ABELL0085 & 1 & 0.15 & 2.86 & 0.414 & 0.017 & 0.013 & FR & Re\\
J004244.68-093316.3 & ABELL0085 & 1 & 0.33 & 3.24 & 0.06 & 0.01 & 0.007 & SR & Re\\
J004310.12-095141.2 & ABELL0085 & 1 & 0.08 & 6.8 & 0.059 & 0.009 & 0.006 & SR & Re\\
J011421.54+001046.9 & ABELL0168 & 1 & 0.08 & 4.73 & 0.087 & 0.007 & 0.005 & FR & Re\\
J011446.94+003128.8 & ABELL0168 & 1 & 0.32 & 3.53 & 0.255 & 0.008 & 0.007 & FR & Re\\
J011454.21+003026.5 & ABELL0168 & 1 & 0.4 & 2.75 & 0.231 & 0.012 & 0.011 & FR & Re\\
J011454.25+001811.8 & ABELL0168 & 1 & 0.18 & 4.36 & 0.304 & 0.007 & 0.006 & FR & Re\\
J011457.59+002550.8 & ABELL0168 & 1 & 0.1 & 10.81 & 0.046 & 0.005 & 0.004 & SR & Re/2\\
J011459.61+001533.1 & ABELL0168 & 1 & 0.37 & 2.62 & 0.335 & 0.013 & 0.01 & FR & Re\\
J011507.33+002756.8 & ABELL0168 & 1 & 0.45 & 4.75 & 0.345 & 0.011 & 0.009 & FR & Re\\
J011508.73+003433.5 & ABELL0168 & 1 & 0.25 & 2.11 & 0.281 & 0.011 & 0.01 & FR & Re\\
J011515.78+001248.4 & ABELL0168 & 1 & 0.05 & 4.27 & 0.073 & 0.008 & 0.006 & FR & Re\\
J011516.77+001108.3 & ABELL0168 & 1 & 0.24 & 3.99 & 0.224 & 0.013 & 0.01 & FR & Re\\
J011531.18+001757.2 & ABELL0168 & 1 & 0.2 & 3.4 & 0.252 & 0.009 & 0.008 & FR & Re\\
J011612.79-000628.3 & ABELL0168 & 1 & 0.03 & 3.66 & 0.166 & 0.009 & 0.008 & FR & Re\\
J215432.20-070924.1 & ABELL2399 & 1 & 0.33 & 3.98 & 0.507 & 0.048 & 0.031 & FR & Re/2\\
J215445.80-072029.2 & ABELL2399 & 1 & 0.32 & 3.25 & 0.289 & 0.012 & 0.009 & FR & Re\\
J215447.94-074329.7 & ABELL2399 & 1 & 0.21 & 1.32 & 0.327 & 0.013 & 0.011 & FR & 2Re\\
J215457.43-073551.3 & ABELL2399 & 0 & 0.12 & 3.76 & 0.053 & 0.011 & 0.008 & SR & Re\\
J215619.00-075515.6 & ABELL2399 & 1 & 0.18 & 2.1 & 0.27 & 0.014 & 0.013 & FR & Re\\
J215624.58-081159.8 & ABELL2399 & 1 & 0.72 & 4.57 & 0.633 & 0.021 & 0.013 & FR & Re\\
J215628.95-074516.1 & ABELL2399 & 1 & 0.01 & 3.63 & 0.187 & 0.029 & 0.018 & FR & Re\\
J215634.45-075217.5 & ABELL2399 & 1 & 0.5 & 2.42 & 0.157 & 0.022 & 0.017 & 2S & Re\\
J215635.58-075616.9 & ABELL2399 & 1 & 0.41 & 3.58 & 0.499 & 0.014 & 0.011 & FR & Re\\
J215637.29-074043.0 & ABELL2399 & 1 & 0.07 & 4.38 & 0.386 & 0.014 & 0.011 & FR & Re\\
J215643.13-073259.8 & ABELL2399 & 1 & 0.2 & 3.64 & 0.055 & 0.013 & 0.009 & SR & Re\\
J215646.76-065650.3 & ABELL2399 & 0 & 0.31 & 4.98 & 0.242 & 0.011 & 0.009 & FR & Re\\
J215650.44-074111.3 & ABELL2399 & 1 & 0.56 & 3.83 & 0.577 & 0.018 & 0.013 & FR & Re\\
J215653.48-075405.5 & ABELL2399 & 1 & 0.51 & 2.13 & 0.394 & 0.021 & 0.019 & FR & Re\\
J215656.92-065751.3 & ABELL2399 & 0 & 0.17 & 2.14 & 0.215 & 0.02 & 0.018 & FR & Re\\
J215658.25-074910.7 & ABELL2399 & 1 & 0.11 & 3.21 & 0.215 & 0.021 & 0.017 & FR & Re\\
J215701.22-075415.2 & ABELL2399 & 1 & 0.33 & 4.1 & 0.073 & 0.015 & 0.01 & SR & Re\\
J215701.35-074653.3 & ABELL2399 & 1 & 0.21 & 2.03 & 0.206 & 0.017 & 0.015 & FR & Re\\
J215701.71-075022.5 & ABELL2399 & 1 & 0.29 & 6.96 & 0.208 & 0.009 & 0.007 & FR & Re\\
J215716.83-075450.5 & ABELL2399 & 1 & 0.29 & 3.3 & 0.307 & 0.015 & 0.014 & FR & Re\\
J215721.41-074846.8 & ABELL2399 & 1 & 0.16 & 4.18 & 0.132 & 0.015 & 0.011 & FR & Re\\
J215723.40-075814.0 & ABELL2399 & 1 & 0.29 & 3.68 & 0.136 & 0.011 & 0.009 & SR & Re\\
J215726.31-075137.7 & ABELL2399 & 1 & 0.28 & 4.57 & 0.405 & 0.028 & 0.02 & FR & Re\\
J215727.30-073357.6 & ABELL2399 & 1 & 0.34 & 3.2 & 0.307 & 0.014 & 0.012 & FR & Re\\
J215727.63-074812.8 & ABELL2399 & 1 & 0.48 & 2.7 & 0.468 & 0.013 & 0.011 & FR & Re\\
\hline
\end{tabular}
\end{table*}

\begin{table*}
\centering
\begin{tabular}{|c|c|c|c|c|c|c|c|c|c|}
\hline
Galaxy Name & Cluster & Membership & Ellipticity & Re & $\lambda_R$ & $\Delta\lambda_R$ & $\Delta\lambda_{R}$ & Kinematic & Fiducial \\
& & & & ($''$ Major Axis) & & & (MC) & Class & Radius\\
\hline
J215729.42-074744.5 & ABELL2399 & 1 & 0.45 & 4.85 & 0.284 & 0.009 & 0.008 & FR & Re\\
J215733.47-074739.2 & ABELL2399 & 1 & 0.33 & 4.67 & 0.21 & 0.008 & 0.006 & FR & Re\\
J215733.72-072729.3 & ABELL2399 & 1 & 0.5 & 4.53 & 0.352 & 0.01 & 0.008 & FR & Re\\
J215743.17-072347.5 & ABELL2399 & 1 & 0.11 & 3.43 & 0.335 & 0.011 & 0.009 & FR & Re\\
J215743.24-074545.1 & ABELL2399 & 1 & 0.2 & 3.93 & 0.135 & 0.014 & 0.011 & SR & Re\\
J215745.05-075701.8 & ABELL2399 & 1 & 0.38 & 1.89 & 0.525 & 0.012 & 0.009 & FR & 2Re\\
J215753.00-074419.0 & ABELL2399 & 1 & 0.06 & 3.87 & 0.289 & 0.015 & 0.013 & FR & Re\\
J215806.62-080642.4 & ABELL2399 & 1 & 0.13 & 2.97 & 0.298 & 0.016 & 0.012 & FR & Re\\
J215807.50-075545.4 & ABELL2399 & 1 & 0.39 & 5.13 & 0.318 & 0.014 & 0.012 & FR & Re\\
J215810.04-074801.4 & ABELL2399 & 1 & 0.4 & 2.74 & 0.486 & 0.016 & 0.015 & FR & Re\\
J215811.35-072654.0 & ABELL2399 & 1 & 0.24 & 1.99 & 0.366 & 0.014 & 0.012 & FR & 2Re\\
J215840.77-074939.8 & ABELL2399 & 0 & 0.32 & 3.03 & 0.307 & 0.023 & 0.018 & FR & Re\\
J215902.71-073930.0 & ABELL2399 & 1 & 0.52 & 3.89 & 0.411 & 0.024 & 0.019 & FR & Re\\
J215945.43-072312.3 & ABELL2399 & 0 & 0.11 & 3.63 & 0.095 & 0.01 & 0.008 & SR & Re\\
\hline
\end{tabular}
\caption{This table presents data for the 79 ETGs in the SAMI Pilot Survey. The columns give the galaxy name, cluster name, cluster membership allocation (1 implies cluster member and 0 implies non-member), the measured ellipticity, the measured effective radius in arcseconds, the analytic measurement of $\lambda_{R}$ and the uncertainty on the measurement, the MC-calculated uncertainty on $\lambda_{R}$, the kinematic classification from the analytic $\lambda_{R}$ measurement and the fiducial radius at which the ellipticity, PA and $\lambda_{R}$ were measured for each galaxy.}
\label{tab:all_gals}
\end{table*}

\end{document}